\documentclass[twocolumn,aps,prb]{revtex4-1} 
\usepackage{dcolumn}
\usepackage{amssymb}
\usepackage{amsmath} 
\usepackage{bm} 
\usepackage{graphicx}
\usepackage{float} 
\usepackage{subfigure} 
\usepackage{stackengine}
\usepackage{hyperref}
\hypersetup{colorlinks,linkcolor={red},citecolor={blue},urlcolor={blue}}
\newcommand{\e}[1]{e^{#1}} 
\newcommand{\ket}[1]{| #1 \rangle} 

\newcommand{\expect}[2]{\langle #1 | #2 | #1 \rangle} 

\begin{document}

\title{Characterizing The Many-Body Localization via Studying State
  Sensitivity to Boundary Conditions}

\author{Mohammad Pouranvari}

\affiliation {Department of Physics, Faculty of Basic Sciences,
  University of Mazandaran, P. O. Box 47416-95447, Babolsar, Iran}

\author{Shiuan-Fan Liou}
\affiliation{Private sector}

\date{\today}

\begin{abstract}
  We introduce novel characterizations for many-body phase transitions
  between delocalized and localized phases based on the system's
  sensitivity to boundary conditions. In particular, we change
  boundary conditions from periodic to antiperiodic and calculate
  shift in the system's energy and shifts in the single-particle
  density matrix eigenvalues in the corresponding energy window. We
  employ the typical model for studying MBL, a one-dimensional
  disordered system of fermions with nearest-neighbor repulsive
  interaction where disorder is introduced as randomness on on-site
  energies. By calculating numerically the shifts in the system's
  energy and eigenvalues of the single-particle density matrix, we
  observe that in the localized regime, both shifts are vanishing;
  while in the extended regime, both shifts are on the order of the
  corresponding level spacing. We also applied these characterizations
  of the phase transition to the case of having next-nearest-neighbor
  interactions in addition to the nearest-neighbor interactions, and
  studied its effect on the transition.
\end{abstract}

\maketitle

\section{Introduction}\label{Introduction}
In a free fermion model with extending plain-wave eigenfunctions,
fermions move freely in the entire system. Introducing randomness in
the free fermion model, which represents disorder, leads to the
localization of fermions due to the quantum interference, proposed by
Anderson\cite{PhysRev.109.1492}. This phenomenon, so-called Anderson
localization, has been widely studied numerically, analytically, and
experimentally. In general, the Anderson localization is associated
with the symmetry and dimension of the
system\cite{RevModPhys.81.865}. For one- and two-dimensional systems,
any infinitesimal uncorrelated randomness makes the system
localized\cite{RevModPhys.80.1355}. For a three-dimensional (3D)
system, however, there exists a non-zero critical disorder strength at
which a quantum phase transition between localized and delocalized
phase occurs\cite{PhysRevLett.47.1546}. In a weak randomness regime,
system is delocalized; as the disorder strength increases and hits the
critical value, the system becomes localized. Besides, a phase
transition from a delocalized to a localized phases can be seen in the
energy resolution, if the system under study has mobility edges. A 3D
Anderson model, for example, has localized phases at both tails of the
energy spectrum, and delocalized phases in the middle of the
spectrum\cite{PhysRevB.5.2931,PhysRevLett.82.4062}. Thus, as the
system's energy changes from one energy window to another, it will
undergo a phase transition between localized and delocalized phases.

Another interesting phenomenon arises when interaction is introduced
in the Anderson model, whence we encounter the following questions:
Does disorder suppress the effect of the interaction? or interaction
effect is so strong that makes the interacting system delocalized?  More
interestingly, what role does temperature play in such a system?  We
can also ask these questions from the perspective of statistical
physics: It is assumed that a classical ergodic system can visit its
whole phase space after a finite time, so the averaging a physical quantity
over time is the same as averaging over the whole phase space. In this
perspective, the question of the ergodicity of a random interacting
system is important\cite{
  doi:10.1146/annurev-conmatphys-031214-014726}. Answering the above
questions is one of the hot research topics. By now, we know that in
the interacting systems, at a non-zero temperature, a phase transition
between localized and delocalized phases arises by varying disorder
strength\cite{Altman2018, RevModPhys.91.021001}. In strong disorder
regime, the conductivity-- even at a non-zero temperature-- is zero,
and the state of the system is localized in Fock space. The phase is
thus called many-body localized (MBL) phase\cite{BASKO20061126,
  doi:10.1002/andp.201600302, ALET2018498,
  doi:10.1146/annurev-conmatphys-031214-014701}. The states in the MBL
phase do not thermalize in the sense that after a long time, its
properties still depend on the initial state of the system (i.e. local
integrals of motion constrain the system); in other words, the system
carries the information of the initial
states\cite{doi:10.1146/annurev-conmatphys-031214-014726}. Thus, an
MBL phase is an out-of-equilibrium phase, and the laws of statistical
mechanics are not obeyed. On the other hand, in weak disorder regime,
a part of the system acts as a bath for the remainder, such that the
eigenstate thermalization hypothesis (ETH)\cite{PhysRevA.43.2046,
  PhysRevE.50.888, Rigol2008, Deutsch_2018} can be applied, and the
system thermalizes. In addition, an ETH-MBL phase transition can be
seen at a fixed disorder strength in the energy resolution, i.e. the
mobility edges can also be seen in the interacting
system\cite{PhysRevB.92.195153, PhysRevLett.115.186601}.

Experimentally, phase transition between the ETH and MBL phases has
been witnessed in many systems such as ultra-cold
atoms\cite{RevModPhys.80.885}, trapped ions systems \cite{Smith2016},
optical lattices\cite{doi:10.1146/annurev-conmatphys-031214-014548,
  RevModPhys.82.1225, Kaufman794, Bordia2017,
  Choi1547,PhysRevLett.116.140401, Schreiber842}, and quantum
information devices\cite{Sapienza1352}.

In the remainder of this section, we first cite models with ETH-MBL
phase transitions. Then we mention some of the previously studied
ETH-MBL phase transition characterizations along with our novel
characterization. The typical model employed to study the ETH-MBL
phase transition is the spinless fermion model with constant
nearest-neighbor (NN) hopping and NN interactions (which is the
Jordan-Wigner transformation of the XXZ model for the spin $1/2$);
disorder is introduced by random on-site energies. Some studies also
introduced random interactions into the system\cite{PAPIC2015714,
  PhysRevB.95.020201, PhysRevB.94.201116}. Having MBL phase in
translation-invariant Hamiltonians and systems with delocalized
single-particle spectrum, also have been
investigated\cite{PhysRevLett.117.240601, PhysRevB.94.201116,
  PhysRevLett.118.266601, PhysRevB.90.165137}. Although disorder is
usually described by random on-site energies, some studies reported
that on-site energies with incommensurate periodicity could trigger
ETH-MBL phase
transitions\cite{PhysRevLett.115.186601,PhysRevLett.115.230401,
  PhysRevB.93.184204}; other systems such as a frustrated spin
chain\cite{2018arXiv180705969C} and a system under strong electric
field\cite{PhysRevLett.122.040606} also exhibit ETH-MBL phase
transitions.

Finding a characterization for the ETH-MBL phase transition is part of
the current research. Entanglement entropy (EE) is one candidate that
shows distinguished behavior in the ETH and MBL
phases\cite{PhysRevLett.113.107204, Geraedts_2017}. To calculate EE,
one divides the system into two subsystems $A$ and $B$; reduced
density matrix of each subsystem $\rho_{A/B}$ is calculated by tracing
over degrees of freedom of other subsystem. EE then can be calculated
as $\text{EE}=-tr \rho_{A/B} \log \rho_{A/B}$ \cite{
  RevModPhys.81.865, LAFLORENCIE20161}. EE follows an area-law
behavior in the MBL phase\cite{PhysRevLett.109.017202,
  PhysRevLett.110.260601}; while it obeys volume law in the ETH phase,
where the reduced density matrix of a subsystem approaches the thermal
density matrix. EE thus fluctuates strongly around the
localization-delocalization transition
point\cite{PhysRevX.7.021013}. In addition, the statistics of the low
energy entanglement spectrum (eigenvalues of the reduced density
matrix) has been used as a characterization of the phase
transition. The entanglement spectrum distribution goes from Gaussian
orthogonal in extended regime to a Poisson distribution in localized
regime\cite{doi:10.1002/andp.201700042}.  Furthermore, people have
used the eigen-energies level spacing\cite{PhysRevB.75.155111} and
level dynamics\cite{PhysRevB.99.224202} as characterizations. Some
other methods, such as machine learning is also used for detecting the
MBL phase transition\cite{0953-8984-30-39-395902}. These were just to
name a few, among many others\cite{PhysRevB.94.201112,
  PhysRevB.93.184204, doi:10.1002/andp.201700042,
  PhysRevLett.113.046806, PhysRevLett.115.186601}.

A recent paper\cite{PhysRevLett.115.046603} studied the
single-particle density matrix to distinguish MBL from the ETH
phase. The single-particle density matrices are constructed by the
eigenstates of the Hamiltonian ($H$) of the system in a target energy
window through
\begin{equation}\label{rho}
  \centering
  \rho_{ij} =\expect{\psi}{c^{\dagger}_i c_j},
\end{equation}
where $i$ and $j$ go from $1$ to $L$ (system size), and
$|\psi \rangle$ is the eigenstate of the Hamiltonian. They studied
eigenvalues $\{n\}$ and eigenfunctions of the density matrix,
$\ket{\phi_k}$. Its eigenvalues, which can be interpreted as
occupations of the orbitals, demonstrate the Fock space localization:
Deep in the delocalized phase, $\{n\}$'s are evenly spaced between $0$
and $1$. While, in the localized phase, they tend to be very close to
either $0$ or $1$. Thus, the difference between two successive
eigenvalues of $\rho$ shows different behavior in delocalized and
localized phases. Moreover, they found that eigenfunctions of the
density matrix, $\ket{\phi_k}$ are extended (localized) in delocalized
(localized) phase\cite{doi:10.1002/andp.201600356,
  SciPostPhys.4.1.002}.

We, in this paper, look at the ETH-MBL phase transition from the
perspective of boundary condition effects on the system, namely, we
change the boundary conditions from periodic to antiperiodic and then
study its effects on the system's energy as well as on the eigenvalues
of the single-particle density matrix ($\rho$) at a given energy
window (see section \ref{model} for more detail). Our work is a
generalization of Ref. [\onlinecite{Edwards_1972}] where Anderson
localization in free fermion models is characterized based on the
response of the system to the change in the boundary conditions. The
response of an interacting system to a local perturbation, on the
other hand, is also investigated in
Refs. [\onlinecite{PhysRevB.96.104201, Khemani2015}] as a
characterization of the MBL phase, which is analogous to our work.

Results of our work, in brief, are as follows. In contrast to the MBL
phase, the system's energy and occupation numbers are sensitive to the
boundary conditions in the ETH phase. In MBL phase, shifts in the
system's energy and occupation numbers are vanishing; however, in ETH
phase, both shifts are on the order of the corresponding level
spacing.  We use these metrics in a previously studied model with NN
interactions that has a known ETH-MBL phase transition. We also apply
these characterizations to a model having both NN and NNN
interactions.

The paper's structure is as follows; We first introduce the model and
explain the numerical method in section \ref{model}. The responses of
the Hamiltonian's eigen-energy and the single-particle density matrix
eigenvalues to the boundary conditions, considering only the NN interaction, will
be presented in sections \ref{E} and \ref{n}, respectively. In sec
\ref{NNN}, we introduce the NNN interaction in the model and consider
its effect. We close with some remarks in section \ref{conclusion}.

\section{Method and Model}\label{model}

We consider spinless fermions confined on a one-dimensional (1D) chain
with the nearest-neighbor (NN) hopping; NN and next-nearest-neighbor
(NNN) repulsive density-density interactions as well as diagonal
disorder. The effective Hamiltonian can be written as:
\begin{equation}\label{H}
  \begin{aligned}
    H &= -t \sum _{i=1} ^{L}  \left( c _{i} ^{\dagger} c _{i+1} + h.c. \right) + \sum _{i = 1} ^{L} \mu _{i} ( n _{i} - \frac{1}{2} ) \\
    + & V _{1} \sum _{i = 1} ^{L} (n _{i} - \frac{1}{2} ) ( n _{i+1} - \frac{1}{2} ) + V _{2}\sum _{i= 1} ^{L}  (n _{i} - \frac{1}{2} ) ( n _{i+2} - \frac{1}{2} ).
  \end{aligned}
\end{equation}

\noindent where $c _{i} ^{\dagger} (c _{i})$ is the fermionic creation
(annihilation) operator, creating (annihilating) a fermion on the site
$i$ and $n _{i} = c ^{\dagger} _{i} c _{i}$ is the number
operator. The first term in the Hamiltonian is the NN hopping with
constant strength $t$, which is used as the energy unit in our
calculations and is set to unity. The randomized on-site energies, as
a disorder representation, are described by $\mu$'s. They follow
uniform distribution within $[-W, W]$, where $W$ is called disorder
strength. The last two terms in the Hamiltonian are the constant
repulsive NN and NNN density-density interactions.

To apply boundary conditions, we set
$ c ^{(\dagger)} _{i + L} = c ^{(\dagger)} _{i}$ for the the periodic
boundary condition (PBC) and
$c ^{(\dagger)} _{i + L} = -c ^{(\dagger)} _{i}$ for the antiperiodic
boundary condition (APBC) where $L$ is the length of the 1D chain.

We first diagonalize the Hamiltonian through exact diagonalization
method, and find its eigenvectors and the corresponding
eigenvalues. We use parameter $\epsilon$ which is defined as
$\epsilon = \frac{(E-E_{0})}{(E_{max}-E_{0})}$, where $E$ is the
target energy, $E_{0}$ is the ground state energy, and $E_{max}$
represents the highest energy in the spectrum; it changes between $0$
and $1$ corresponding to the ground state and highest energy,
respectively. We focus on a certain energy window of the spectrum: For
a given $\epsilon$, we calculate the target energy $E$ and select six
eigenstates of $H$ with the energy closest to $E$. For each of these
six eigenstates, we build up the single-particle density matrix $\rho$
from Eq. (\ref{rho}).  By changing the boundary conditions from PBC to
APBC, we calculate the energy shift for each eigenstate:
\begin{equation}
  \delta E_i = | E _{i, PBC} - E _{i, APBC} |,
\end{equation}
where $E _{i}$ is the energy of the $i$th eigenstate; in this way,
$i$th level of the Hamiltonian with PBC is compared with the $i$th
level of the Hamiltonian with APBC. We then take typical averaging
over six eigenstates and take typical disorder average to obtain
$\delta E^{typ}$ (typical average of random variable $x$ is
$\e{\langle \ln x \rangle}$, where $\langle \cdots\rangle$ stands for
arithmetic mean).  In the same manner, we calculate shifts in the
eigenvalues of the single-particle density matrix:
\begin{equation}
  \delta n_i^{(j)} = | n_{i, PBC}^{(j)} - n_{i, APBC}^{(j)} |.
\end{equation}
A typical average on all eigenvalues of the $\rho$ ($j$ goes from
$1$ to $L$), another typical average over the six samples, and a
typical disorder average will be calculated to obtain
$\delta n^{typ}$.

\section{Effect of the boundary change on the
  energy}\label{E}
In a free fermion model, the state of the system is the Slater
determinant of the occupied single-particle eigenstates. In the
localized phase, occupied eigenstates of the system are confined in a
small region of space, while in the delocalized phase, they spread
over entire system. In Ref. [\onlinecite{Edwards_1972}], the effect of
the change in the boundary conditions on the single-particle
eigen-energies of a free fermion model is studied. By changing the
boundary conditions in the localized phase, the single-particle energy
does not change. On the other hand, in the delocalized phase, where
eigenstates of the system are extended, any changes in the boundary
conditions can be seen by the wave-function; these changes are then
reflected in the corresponding energy. Accordingly, the energy shift
for each level, $\delta E$, divided by the average level spacing
$\Delta E$ known as Thouless conductance, is a characterization of the
Anderson phase transition between delocalized and localized phases:
\begin{equation}\label{shiftE}
  \centering
  g_E=\delta E / \Delta E.
\end{equation}

\begin{figure}
  \centering
  \includegraphics[width=0.50\textwidth]{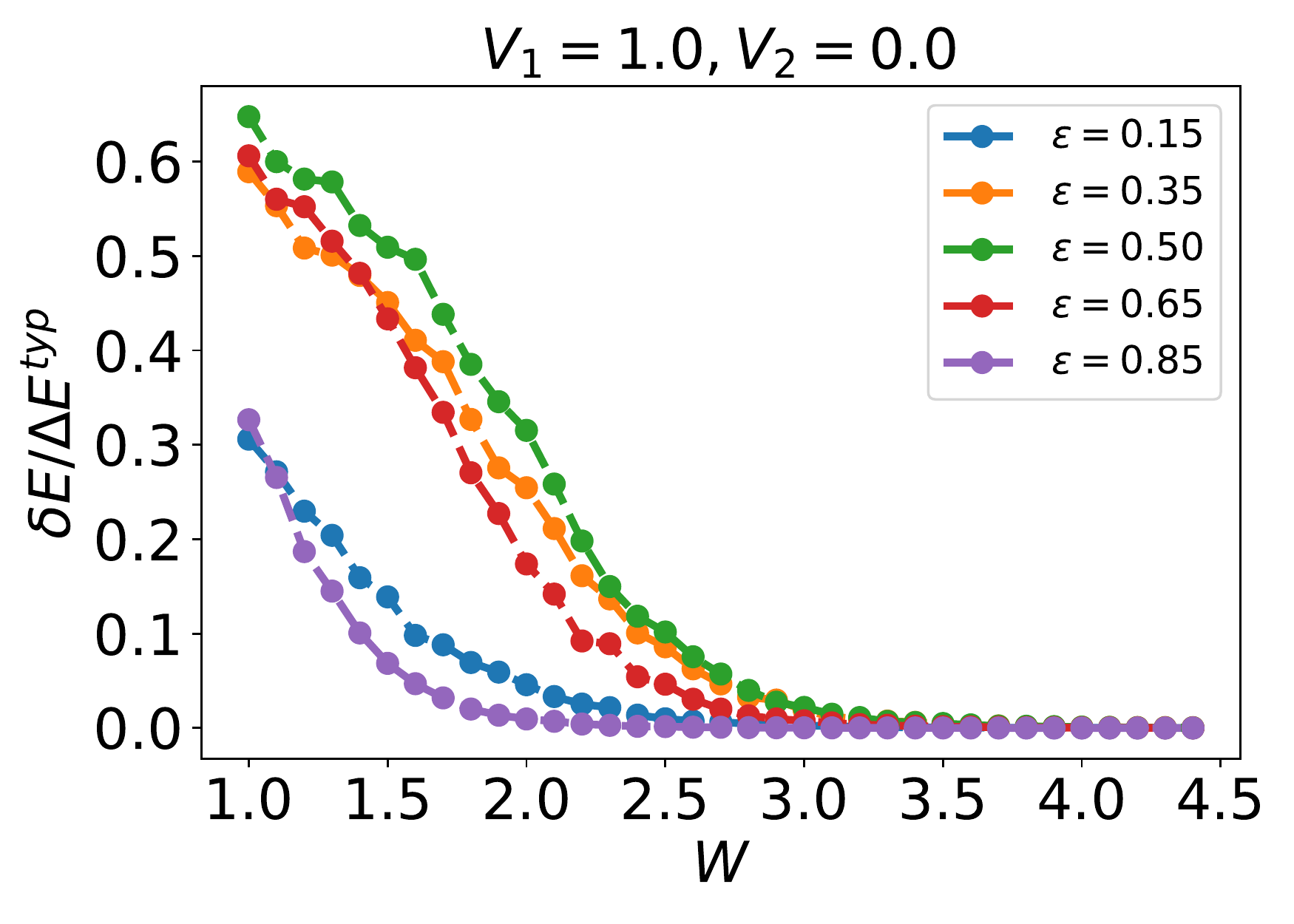}
  \caption{(color online) typical averaged
    $g_E=\frac{\delta E}{ \Delta E}$ for the case of having only NN
    interaction ($V_1=1, V_2=0$) for some selected $\epsilon$ as
    disorder strength $W$ varies. We set $L=14, N=7$. We take typical
    disorder average over altogether $2000$ samples for each data
    point.\label{fig:g_E_eps10}}
\end{figure}

\begin{figure}
  \centering
  \includegraphics[width=0.5\textwidth]{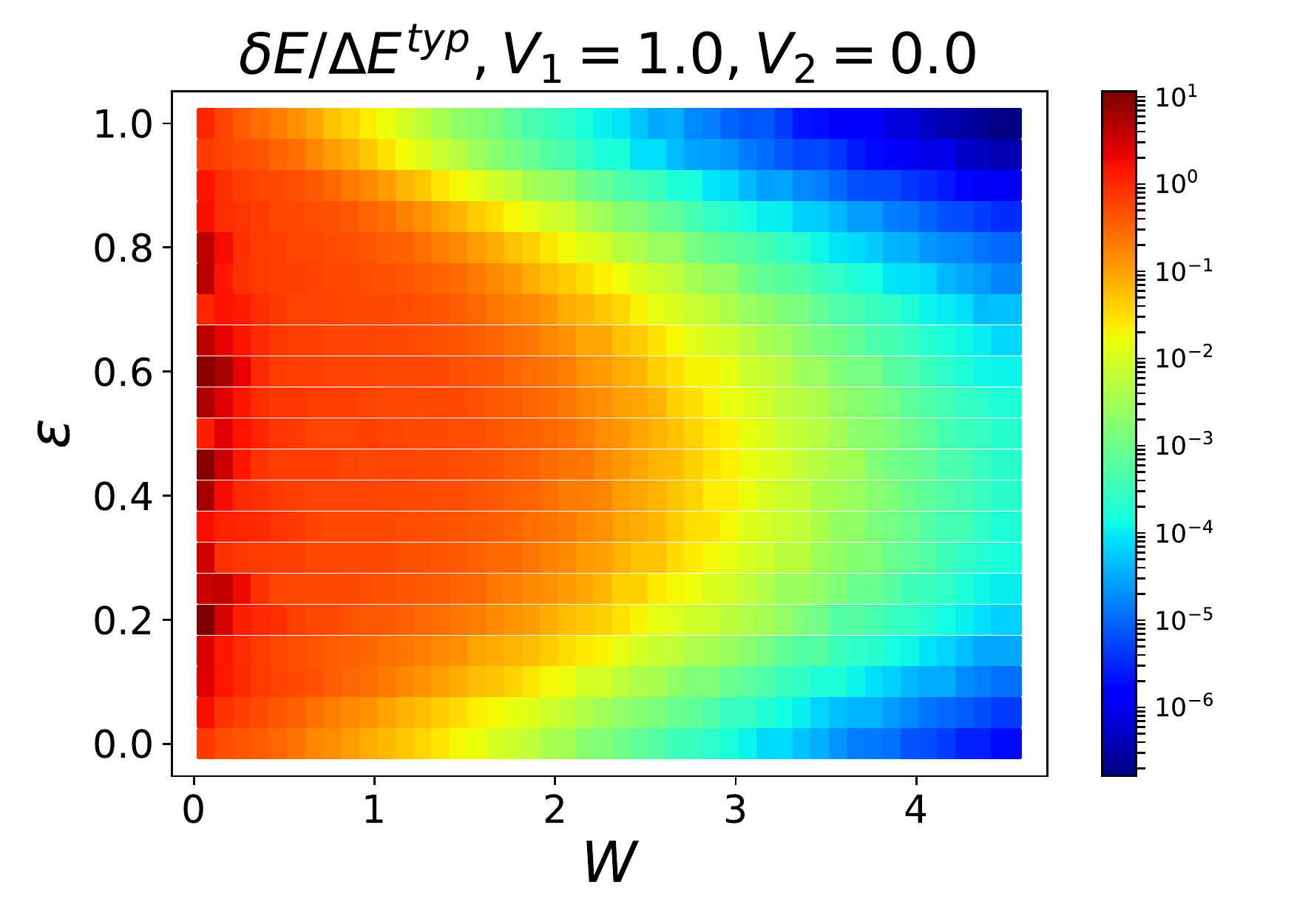}
  \caption{(color online) typical average of
    $g_E=\frac{\delta E}{ \Delta E}$ for the entire spectrum of the
    energy as we change the disorder strength $W$ for the NN
    interaction case ($V_1=1, V_2=0$). System size $L=14, N=7$. We
    take typical disorder average over altogether $2000$ samples for
    each data point.\label{g_E10}}
\end{figure}

We conjecture that if we change boundary conditions for an interacting
system, a similar quantity as Thouless conductance (now for the
system's energy rather than the single-particle eigen-energy) can be
used to characterize the phase transition. In particular, we change
the condition from periodic to antiperiodic (as explained in section
\ref{model}) and calculate $g_E$ for the system's energy. In
Fig. \ref{fig:g_E_eps10}, typical averaged $g_E$ is plotted for some
selected values of the energy for the case of NN interaction of
Eq. (\ref{H}) corresponding to $V_1=1,V_2=0$ (standard deviation of
$g_E$ is plotted in Fig. \ref{fig:sigma}).  We see that deep in the
delocalized phase, the shift in the system's energy is on the order of
level spacing, while deep in the localized phase, the shift is
zero. Based on this plot, in the middle of the spectrum
($\epsilon=0.5$), $g_E$ goes to zero at $W \approx 3.5$, consistent
with the previously obtained results\cite{PhysRevB.91.081103,
  PhysRevB.96.104201, PhysRevB.96.014204, PhysRevB.97.104406}. Also,
$g_E$ is plotted for the whole spectrum of energy in Fig. \ref{g_E10}
as we vary the disorder strength $W$. This plot is also consistent
with the previously obtained results.

\section{Effect of the boundary change on the single-particle density
  matrix}\label{n}

Now, we focus on the effect of boundary conditions on the occupation
numbers of density matrix Eq. (\ref{rho}). First, let us look at the
case of free fermions, where we can write the Hamiltonian of the
system as:
\begin{equation}\label{Hfree}
  \centering
  H_{\text{Free fermion}} = \sum_{i,j=1}^{L}h_{ij}\ c^{\dagger}_i c_j,
\end{equation}
we observe that eigen-functions of the single-particle density matrix
and matrix $h$ are the same (both can be diagonalized by the same
unitary matrix):

\begin{eqnarray}
  \centering
  h_{ij} &=& \sum_k U_{ik} \ \epsilon_{k} \  U_{kj}^{\dagger},\\
  \rho_{ij} &=& \sum_k U_{ik} \ n_{k} \  U_{kj}^{\dagger},
\end{eqnarray}

where $\epsilon_k$ is the single-particle eigen-energy of the
Hamiltonian. By the argument of Thouless\cite{Edwards_1972} that
single-particle eigenstates of the Hamiltonian are sensitive
(insensitive) to the boundary conditions in delocalized (localized)
phase, we can say that eigenvalues of the $\rho$ are sensitive
(insensitive) to the boundary conditions in delocalized (localized)
phase. Thus, we can identify the shifts in the eigenvalues of the
$\rho$ when we change the boundary condition from periodic to
antiperiodic as a probe of the phase transition.

This idea has been verified indirectly before: We know that for a free fermion 
system divided into two subsystems, reduced density matrix of each
subsystem can be written as $\exp(-H_{\text{ent}})$, where
$H_{\text{ent}}$ is called entanglement Hamiltonian and can be
obtained from single-particle density matrix of the corresponding
subsystem\cite{0305-4470-36-14-101}. Effect
of the boundary condition changes on the entanglement Hamiltonian for
free fermion models was studied in Ref
[\onlinecite{PhysRevB.96.045123}]: Boundary condition is changed from
periodic to antiperiodic and shifts in the eigenvalues of the
entanglement Hamiltonian (and thus on the entanglement entropy) are
calculated; and it is shown that they can be used as characterization
of the localized-delocalized phase transition.

For the interacting case, we know that the single-particle density
matrix eigenstates are localized (delocalized) in the localized
(delocalized) phase\cite{PhysRevLett.115.046603}; Thus, we put one
step forwards and conjecture that ETH phase can be distinguished from
the MBL phase by analyzing the shifts of the occupation numbers when
we change boundary conditions. In particular, we change the boundary
condition from periodic to antiperiodic (as described in Section
\ref{model}) and calculate the shifts in the occupation numbers of
single-particle density matrix $\delta n$.

\begin{figure}
  \centering
  \begin{subfigure}{}%
    \includegraphics[width=0.23\textwidth]{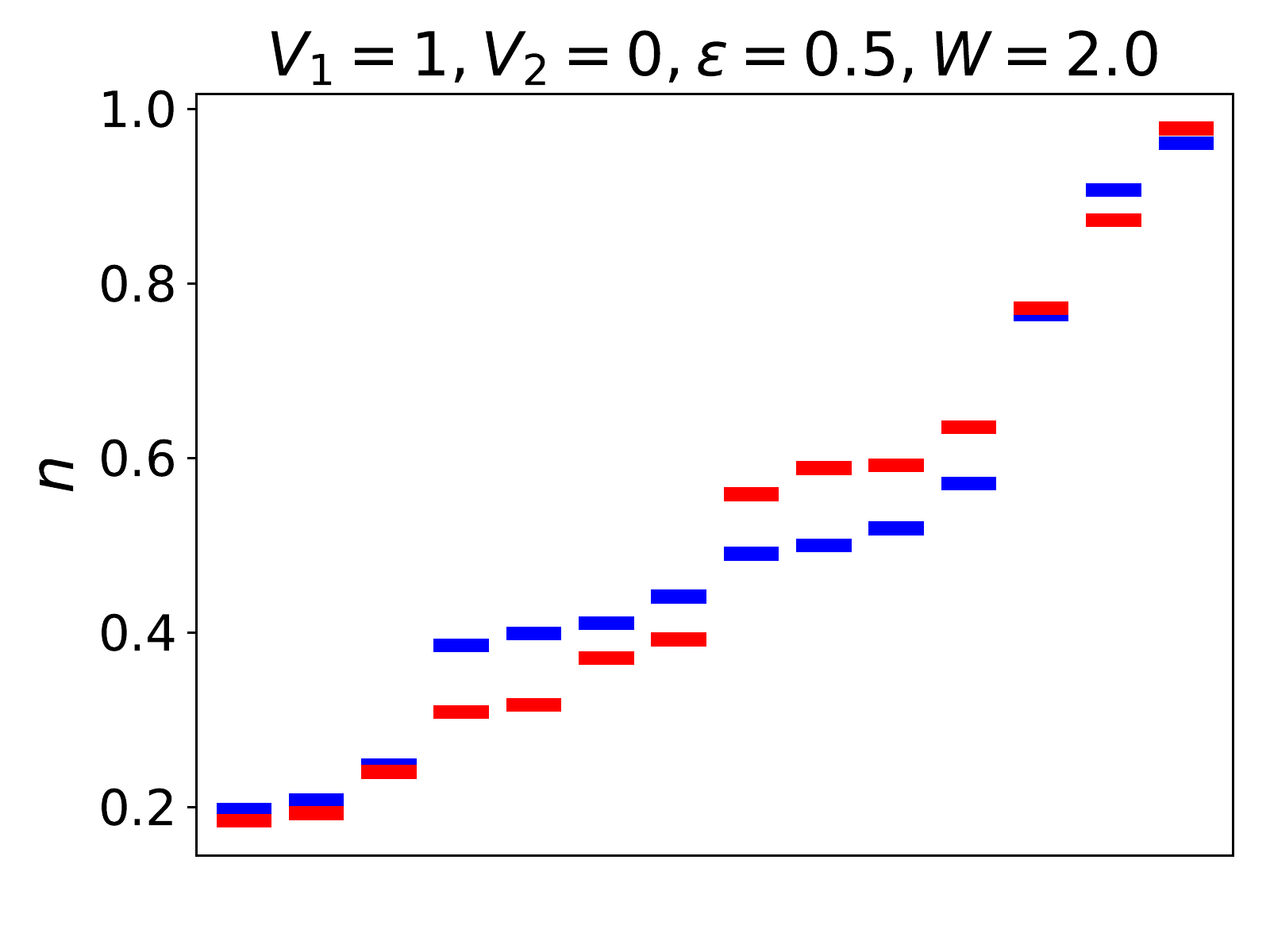}
  \end{subfigure}%
  ~%
  \begin{subfigure}{}%
    \includegraphics[width=0.23\textwidth]{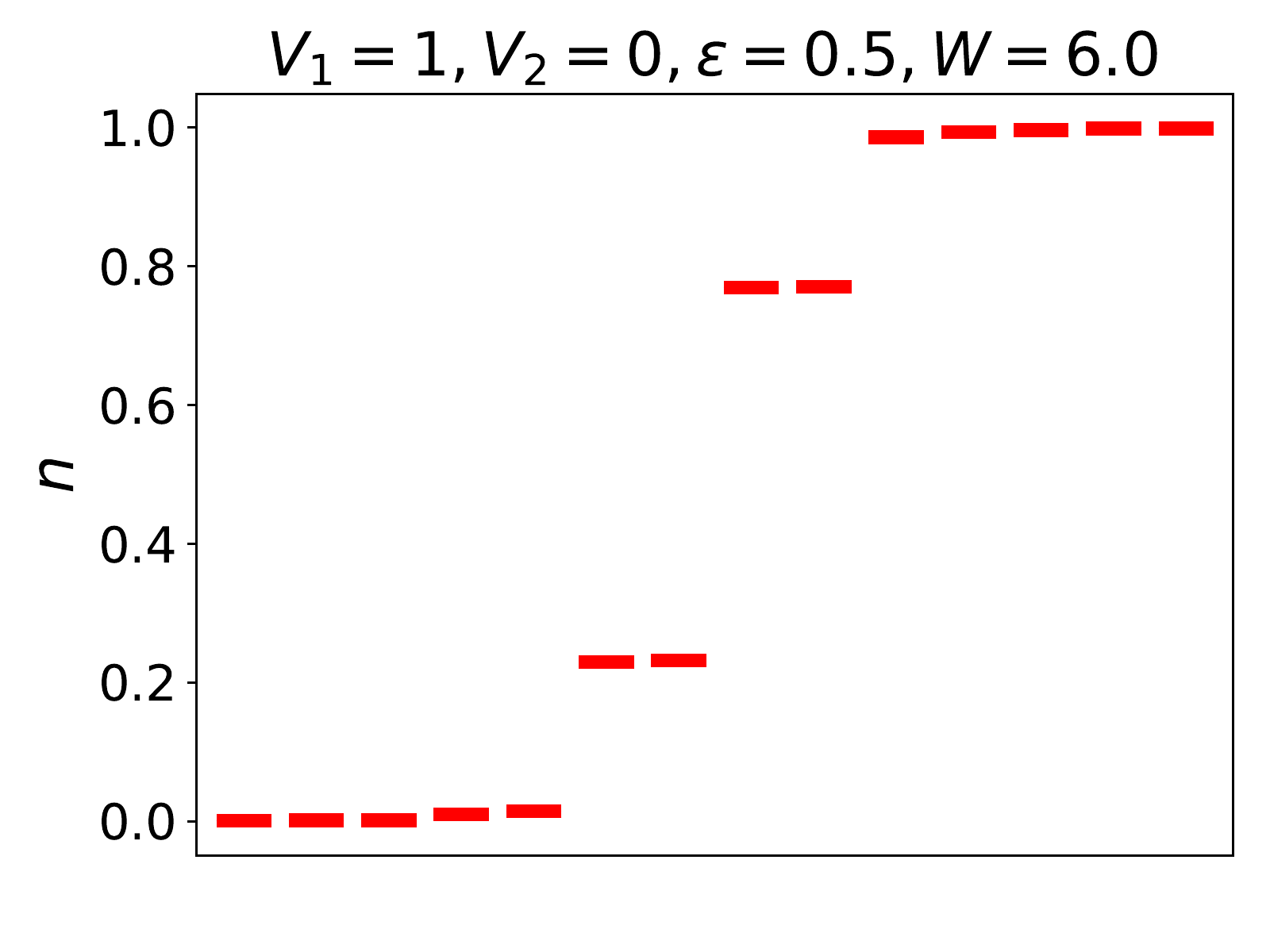}
  \end{subfigure}

  \begin{subfigure}{}%
    \includegraphics[width=0.23\textwidth]{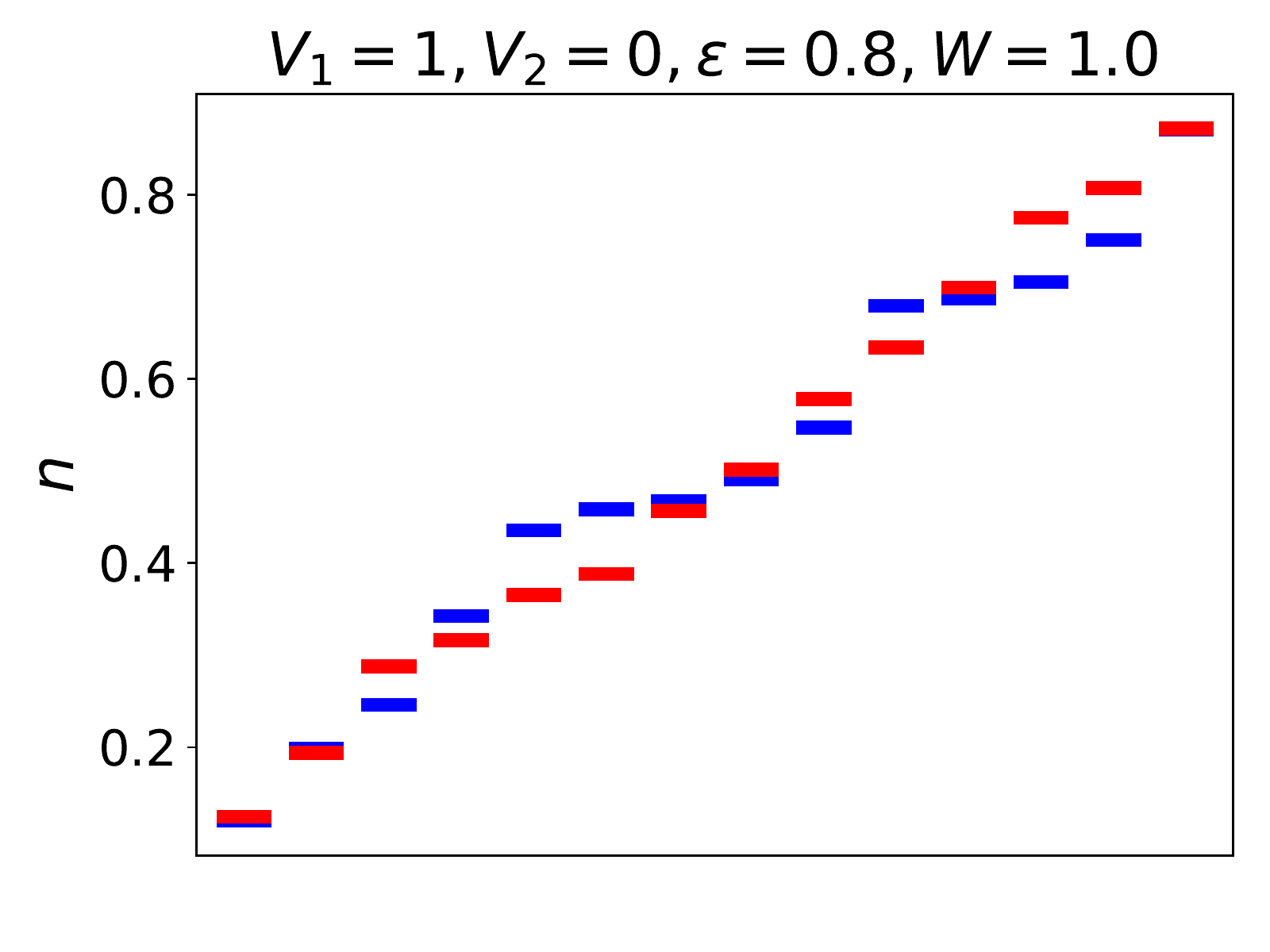}
  \end{subfigure}%
  ~%
  \begin{subfigure}{}%
    \includegraphics[width=0.23\textwidth]{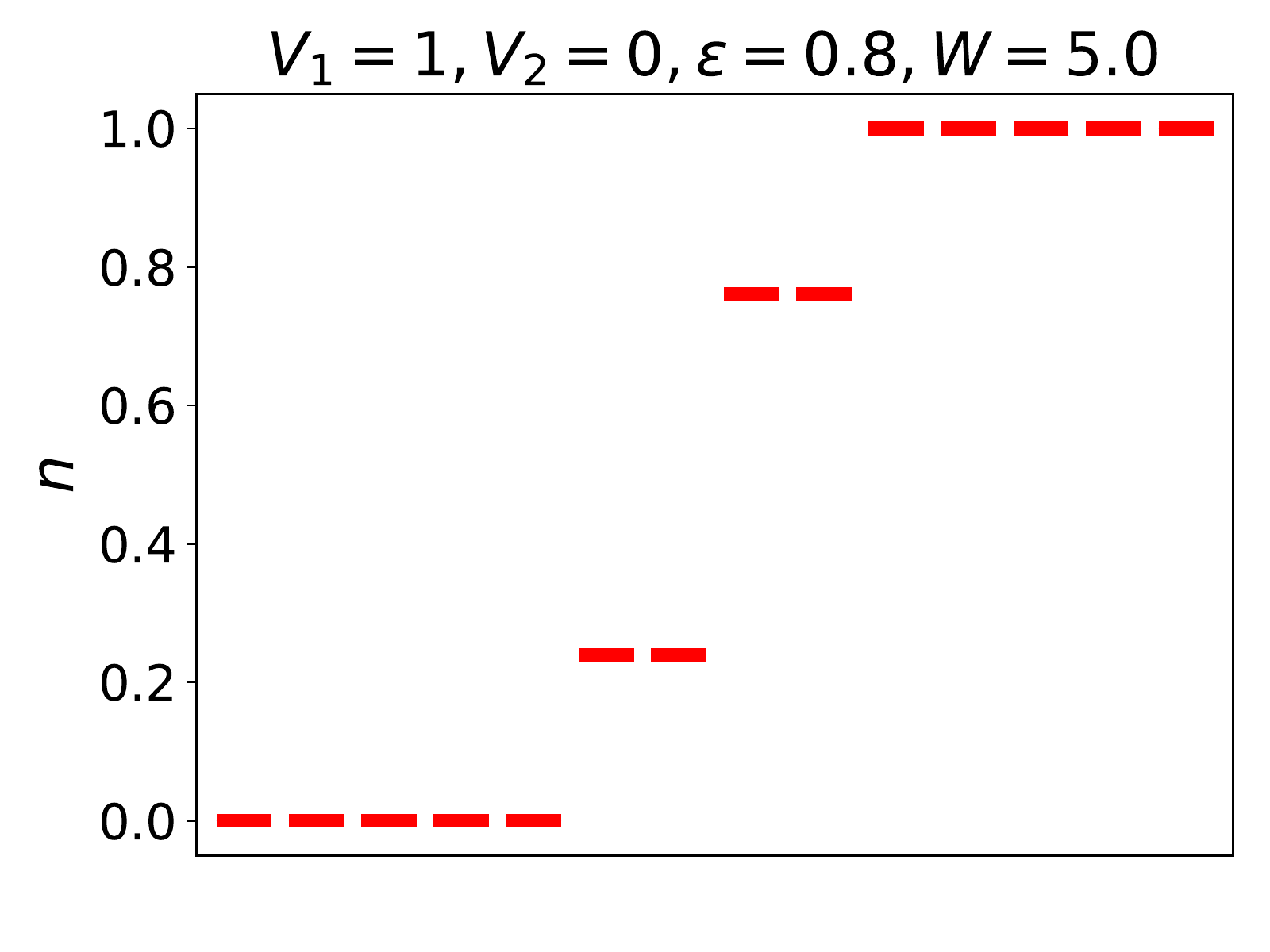}
  \end{subfigure}
  \caption{ (color online) eigenvalues of the single-particle density
    matrix (occupation numbers $\{n\}$) corresponding to periodic
    boundary condition (Blue), and antiperiodic boundary condition
    (red) for NN interaction of Eq.(\ref{H}) ($V_1=1,V_2=0$). In left
    plots, $\epsilon$ and $W$ are chosen such that we are in the
    extended phase, while in the right plots, they correspond to the
    MBL phase. We set $L=14, N=7$. Only one sample is considered, and
    we do not take disorder average. We see that shifts in the
    occupation numbers in the MBL phase are almost zero, while the
    shifts are appreciable in the extended phase. \label{fig:deln10}}
\end{figure}

In Fig. \ref{fig:deln10} we plot occupation numbers for the NN
interaction case of Eq. (\ref{H}) corresponding to $V_1=1,V_2=0$, for
periodic and antiperiodic boundary conditions in extended and MBL
phases. Here, just one sample is considered without disorder
averaging. We can see that in the MBL phase, occupation numbers
corresponding to PBC and APBC are almost identical, and the shifts are
negligible; in contrast, we get a non-vanishing change of the
occupation numbers in the extended phase.

To have a characterization independent of the system size, we divide
$\delta n$ to average level spacing for occupation numbers,
$\Delta n$, and introduce the following as an ETH-MBL phase transition
characterization:
\begin{equation}\label{gn}
  \centering
  g_{n}=\delta n / \Delta n.
\end{equation}

\begin{figure}
  \centering
  \includegraphics[width=0.50\textwidth]{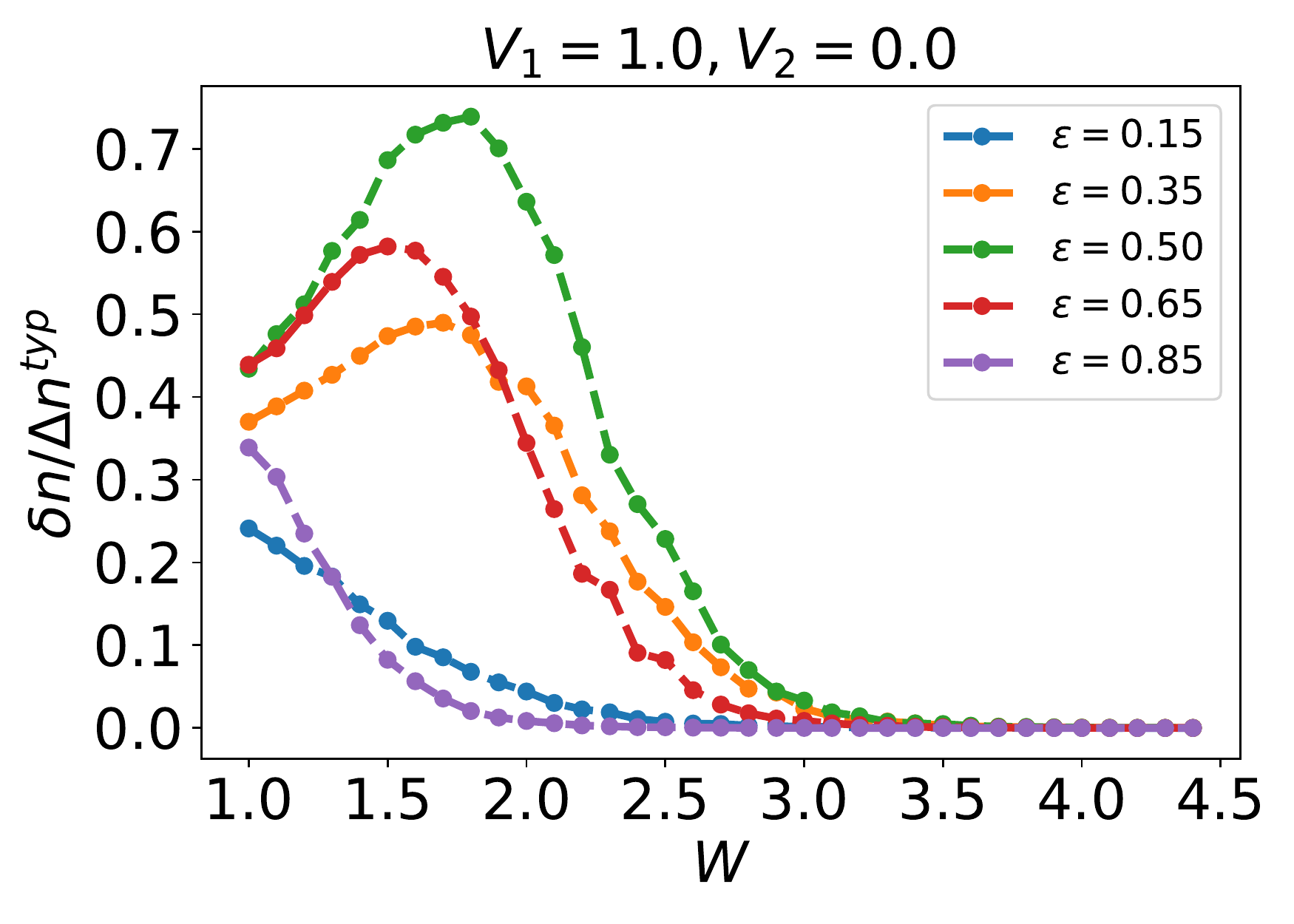}
  \caption{(color online) typical averaged
    $g_n=\frac{\delta n}{ \Delta n}$ for the NN interaction case of
    Eq. (\ref{H}) corresponding to $V_1=1, V_2=0$, for some selected
    values of $\epsilon$'s, as we change the disorder strength $W$. We
    set $L=14, N=7$. We take typical average over altogether $2000$
    samples for each data point.\label{fig:g_n_eps10}}
\end{figure}

We plot typical averaged $g_n$ for the NN case of Eq. (\ref{H})
($V_1=1, V_2=0$) for some selected values of $\epsilon$, as we change
disorder strength $W$ in Fig. \ref{fig:g_n_eps10}. We see that deep
in the delocalized phase, shifts in the eigenvalues of density matrix are on the order of level spacing of the eigenvalues,  and it
vanishes in the localized phase. At the middle of the spectrum
($\epsilon=0.5$), we obtain $W_c \approx 3.6$, consistent with the
previously obtained results\cite{PhysRevB.91.081103,
  PhysRevB.96.104201, PhysRevB.96.014204, PhysRevB.97.104406}. We see
that $g_n$ is not symmetric about the middle of the spectrum and it is
tilted toward smaller $\epsilon$.

\begin{figure} \centering
  \includegraphics[width=0.5\textwidth]{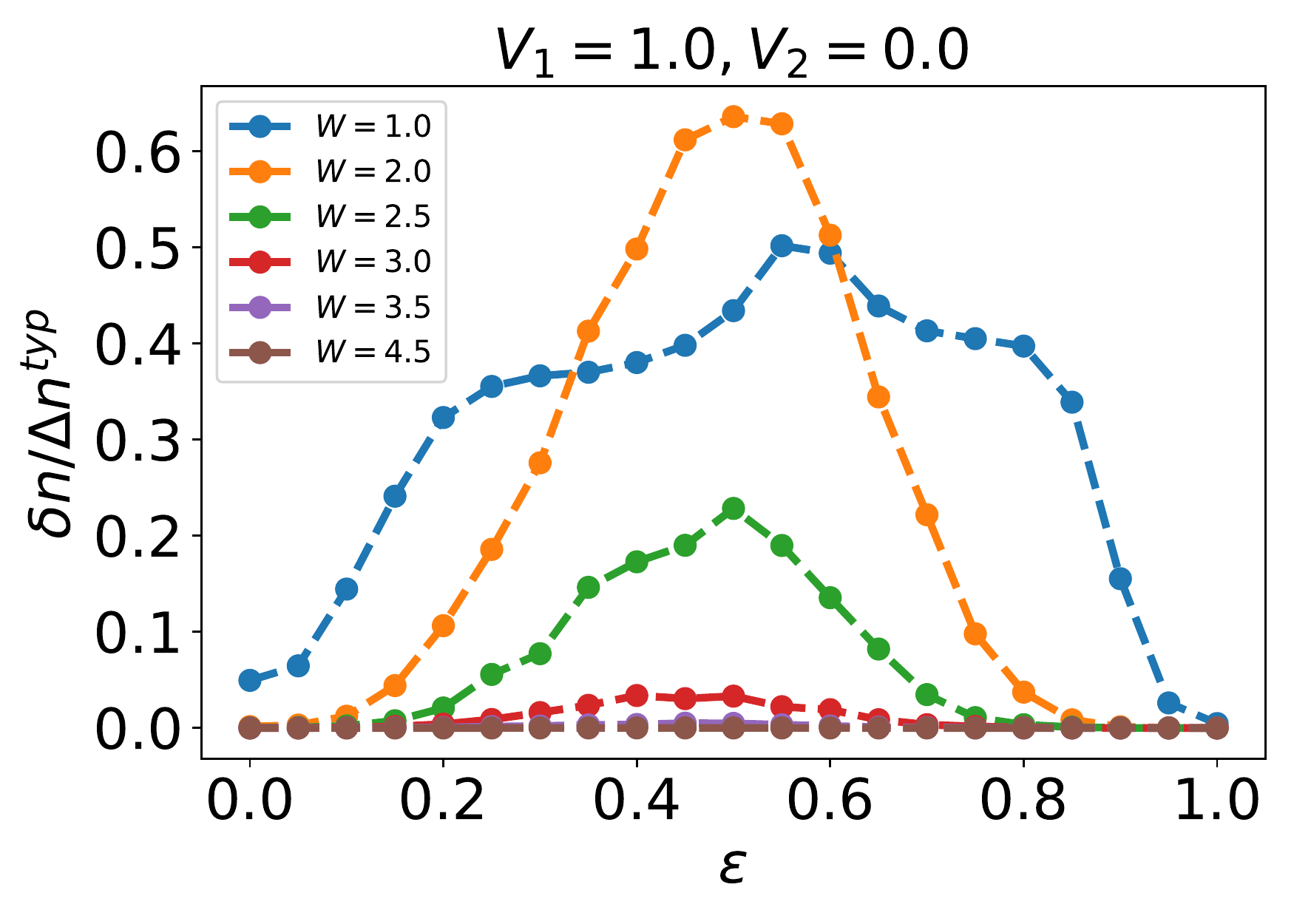}
  \caption{(color online) typical averaged
    $g_n=\frac{\delta n}{ \Delta n}$ for the NN interaction case of
    Eq. (\ref{H}) corresponding to $V_1=1, V_2=0$, for some selected
    values of disorder strength $W$, as energy varies, where we can
    see the mobility edges. We set $L=14, N=7$. We take typical
    average over altogether $2000$ samples for each data
    point.\label{fig:g_n_W10}}
\end{figure}

\begin{figure}
  \centering
  \includegraphics[width=0.50\textwidth]{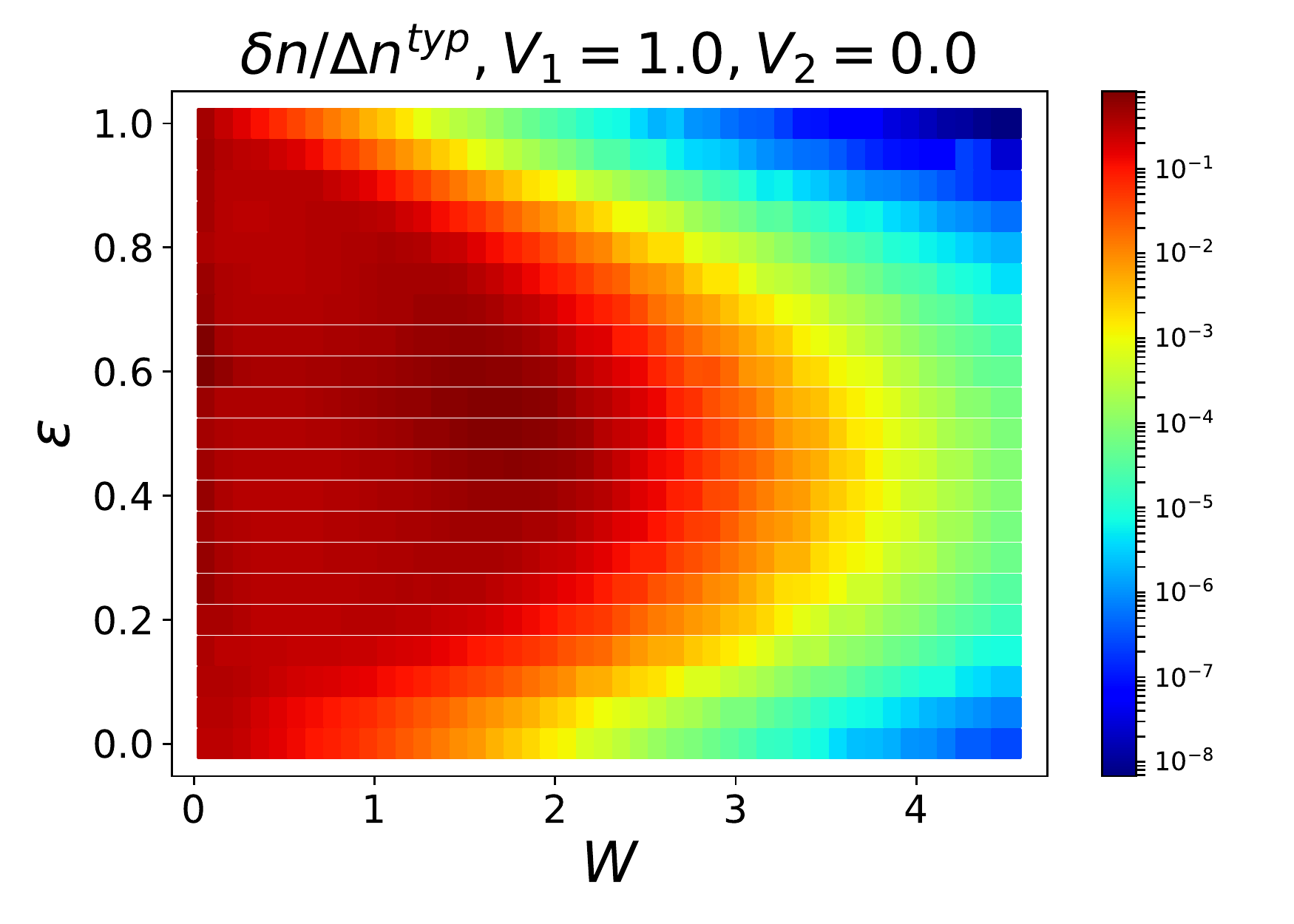}
  \caption{(color online) typical averaged
    $g_n=\frac{\delta n}{ \Delta n}$ for the NN interaction case of
    Eq. (\ref{H}) corresponding to $V_1=1, V_2=0$, for the entire
    energy spectrum as we change the disorder strength $W$. We set
    $L=14, N=7$. We take typical average over altogether $2000$
    samples for each data point.\label{fig:g_n_10}}
\end{figure}

\begin{figure}
  \begin{subfigure}{}%
    \includegraphics[width=0.45\textwidth]{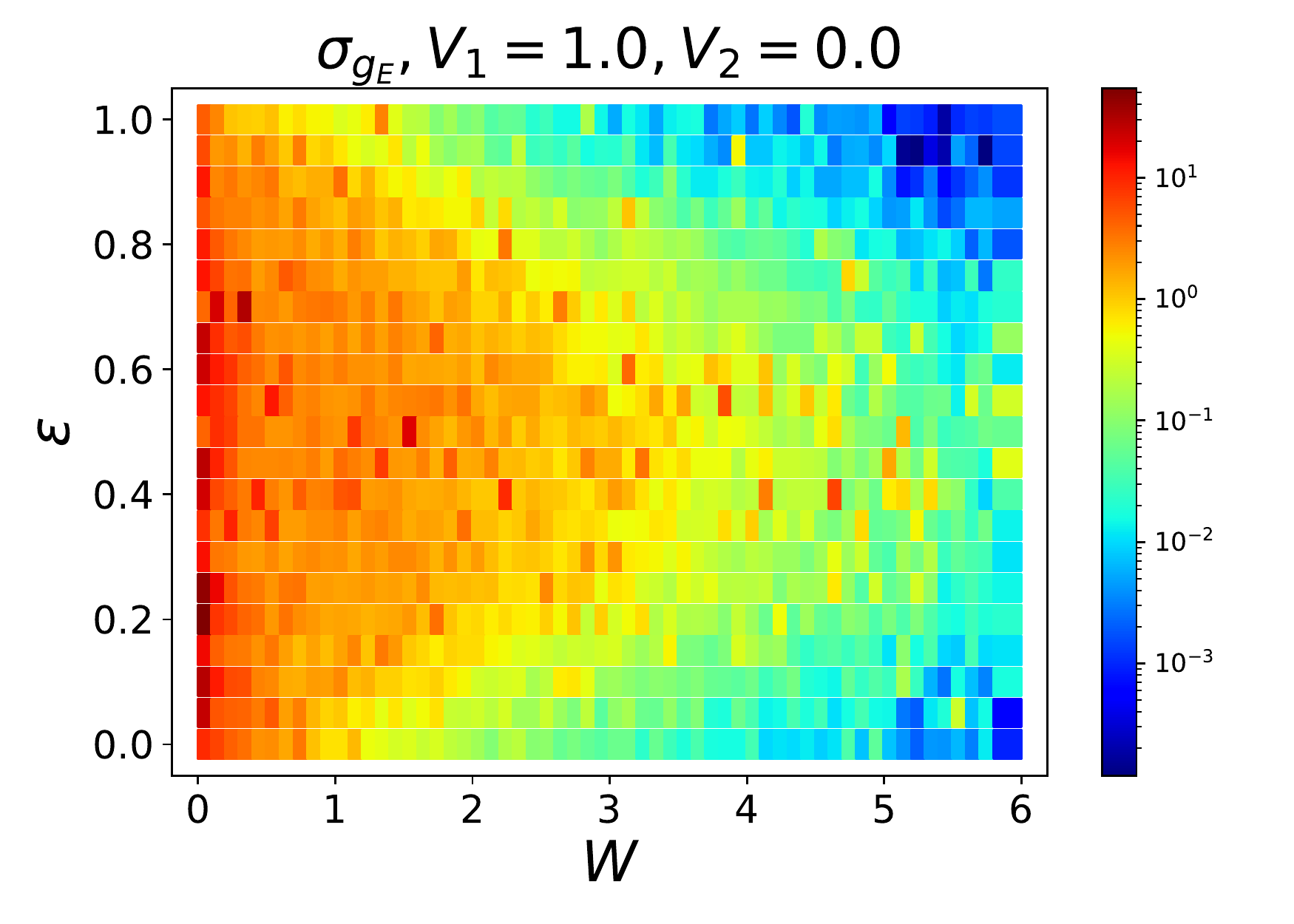}
  \end{subfigure}%
  ~%
  \begin{subfigure}{}%
    \includegraphics[width=0.45\textwidth]{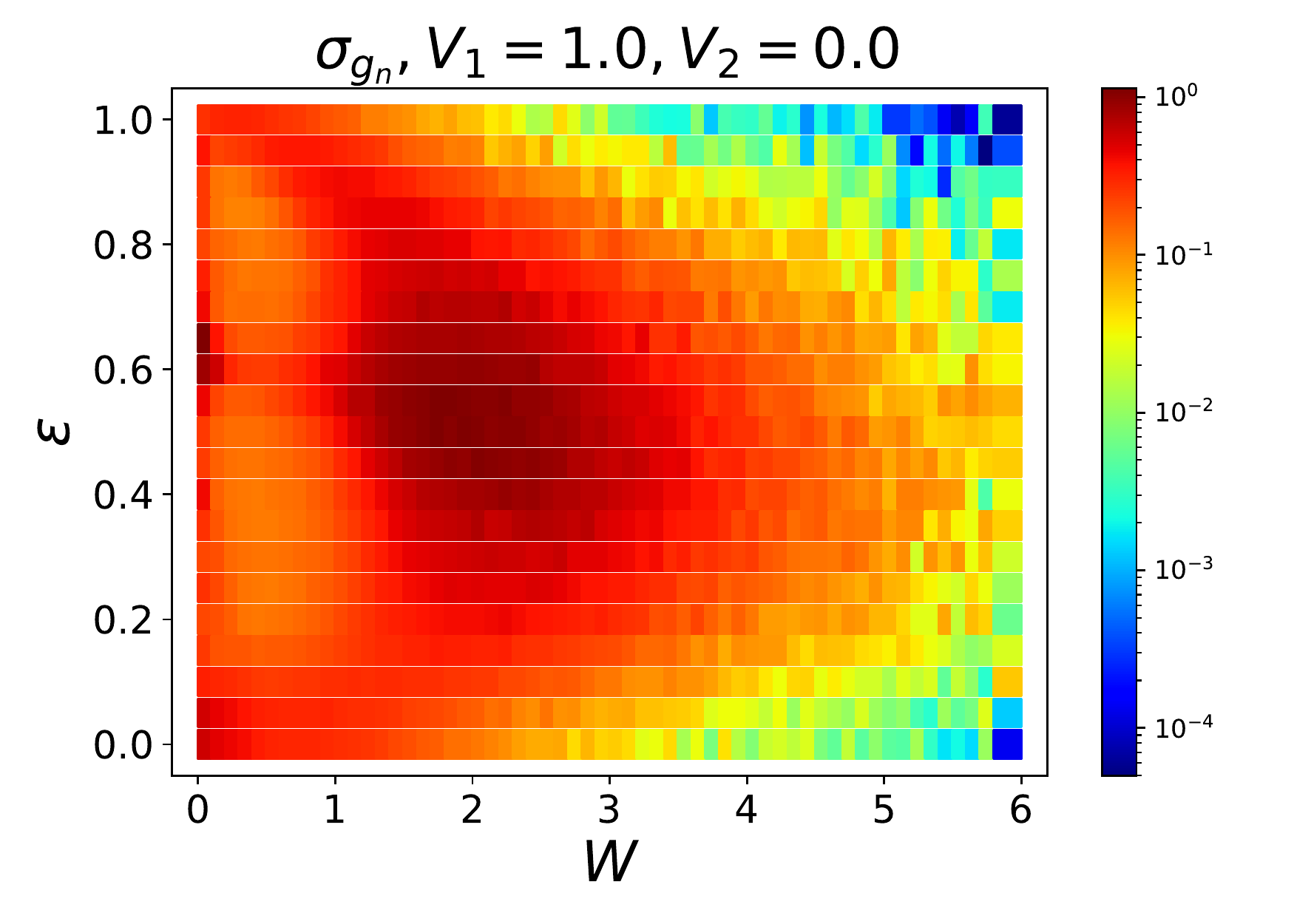}
  \end{subfigure}%
  \caption{(color online) standard deviation of $g_E$ (upper panel)
    and $g_n$ (lower panel) for the entire spectrum of the energy as
    we change the disorder strength $W$ for the NN interaction case
    ($V_1=1, V_2=0$). System size $L=14, N=7$. Number of samples is
    $2000$ for each data point.
    \label{fig:sigma}}
\end{figure}

By looking at $g_n$, we can locate mobility edges, the points in the
energy spectrum, for a fixed value of disorder strength, where phase
changes between delocalized and localized . We calculate $g_n$ for a
fixed value of $W$, as we change $\epsilon$. The results are plotted
in Fig. \ref{fig:g_n_W10}.  As we can see, there are no mobility edges
deep in the localized phase and deep in the delocalized phase,
i.e. for $W=1.0$ and $W=4.5$. For $W=1.0$, $g_n$ is always non-zero,
while for $W=4.5$ it vanishes for all values of $\epsilon$. For other
disorder strength values, we can see mobility edges where $g_n$ goes
to zero. All this information can be summarized in
Fig. \ref{fig:g_n_10}, where $g_n$ is calculated for the entire energy
spectrum as we change disorder strength $W$(standard deviation of
$g_n$ is plotted in Fig. \ref{fig:sigma}).

\section{Nearest-neighbor and next-nearest-neighbor
  interactions}\label{NNN}
\begin{figure*}
  \begin{subfigure}{}%
    \includegraphics[width=0.32\textwidth]{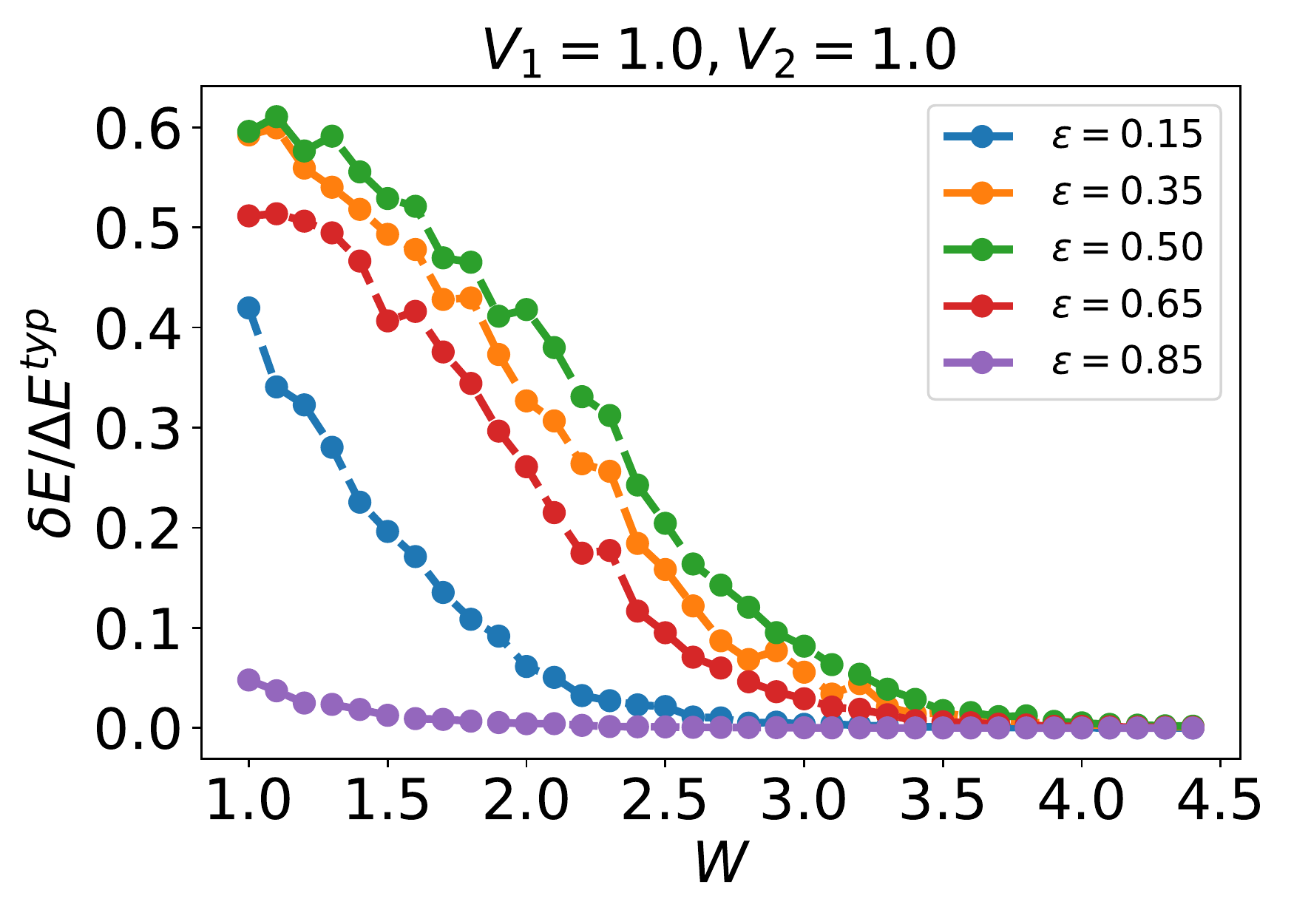}
  \end{subfigure}%
  ~%
  \begin{subfigure}{}%
    \includegraphics[width=0.32\textwidth]{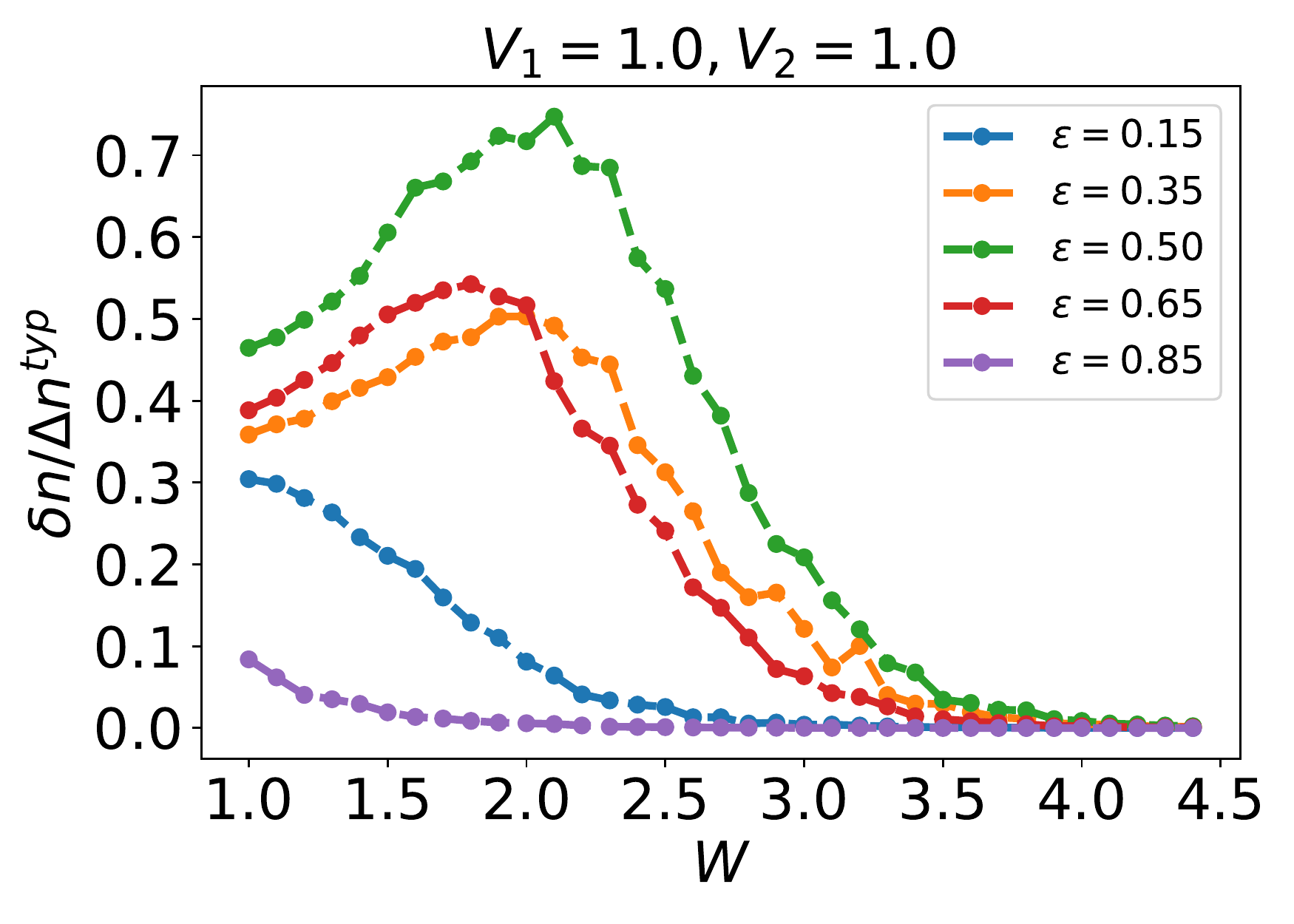}
  \end{subfigure}%
  ~%
  \begin{subfigure}{}%
    \includegraphics[width=0.32\textwidth]{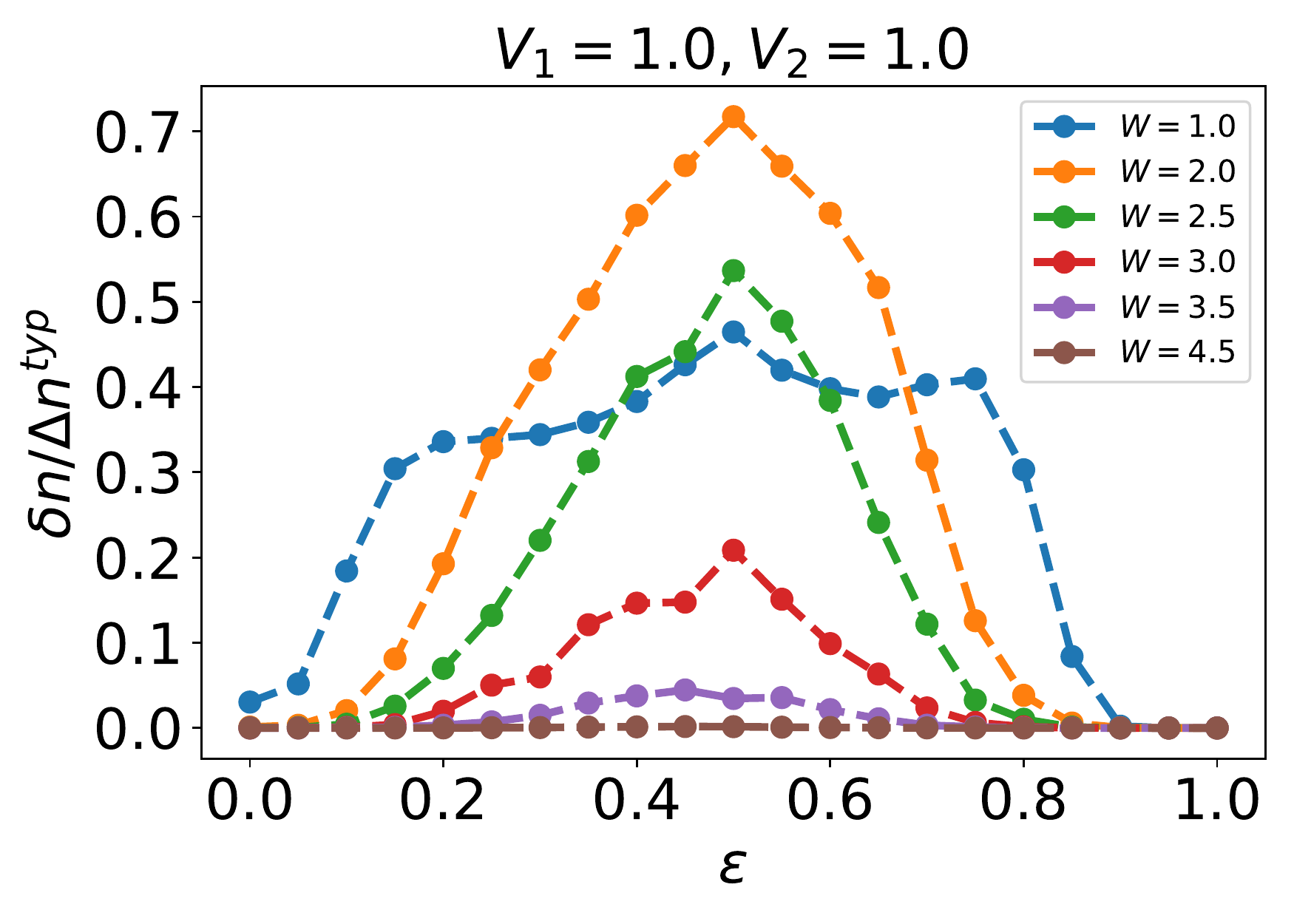}
  \end{subfigure}
  \caption{(color online) typical averaged
    $g_E=\frac{\delta E}{ \Delta E}$ (left panel) and
    $g_n=\frac{\delta n}{ \Delta n}$ (right panel) for the case of
    having both NN and NNN interactions corresponding to
    $V_1=1, V_2=1$ in Eq. (\ref{H}) for some specific $\epsilon$'s as
    disorder strength $W$ varies. Right panel: typical averaged
    $g_n=\frac{\delta n}{ \Delta n}$ for the case of NN and NNN
    interactions for some selected values of disorder strength $W$ as
    energy varies, where we can see the mobility edges. For all plots,
    We set $L=14, N=7$. We take typical average over altogether $2000$
    samples for each data point. \label{fig:shift_n_nnn}}
\end{figure*}

\begin{figure}
  \centering
  \begin{subfigure}{}%
    \includegraphics[width=0.45\textwidth]{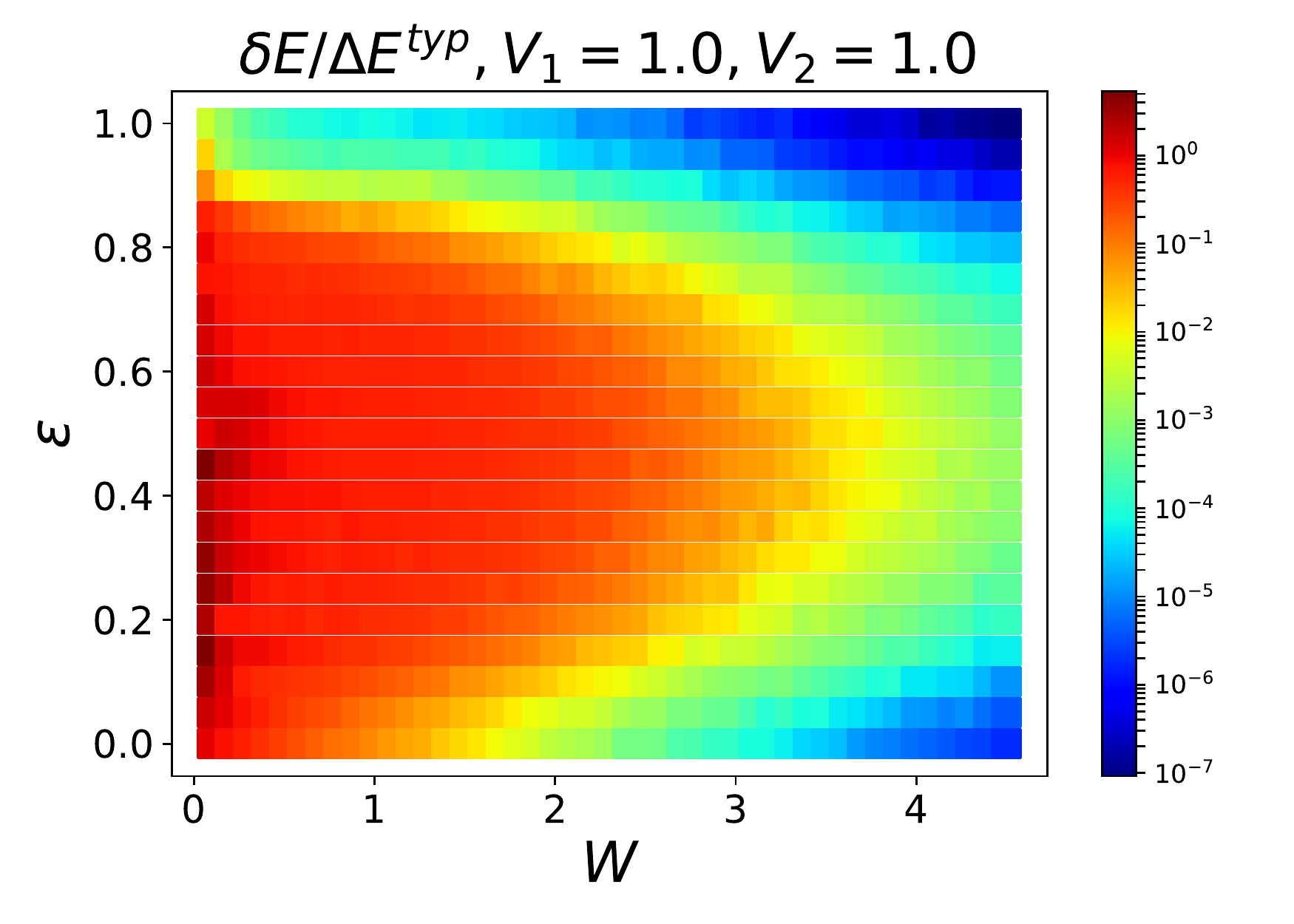}
  \end{subfigure}%
  ~%
  \begin{subfigure}{}%
    \includegraphics[width=0.45\textwidth]{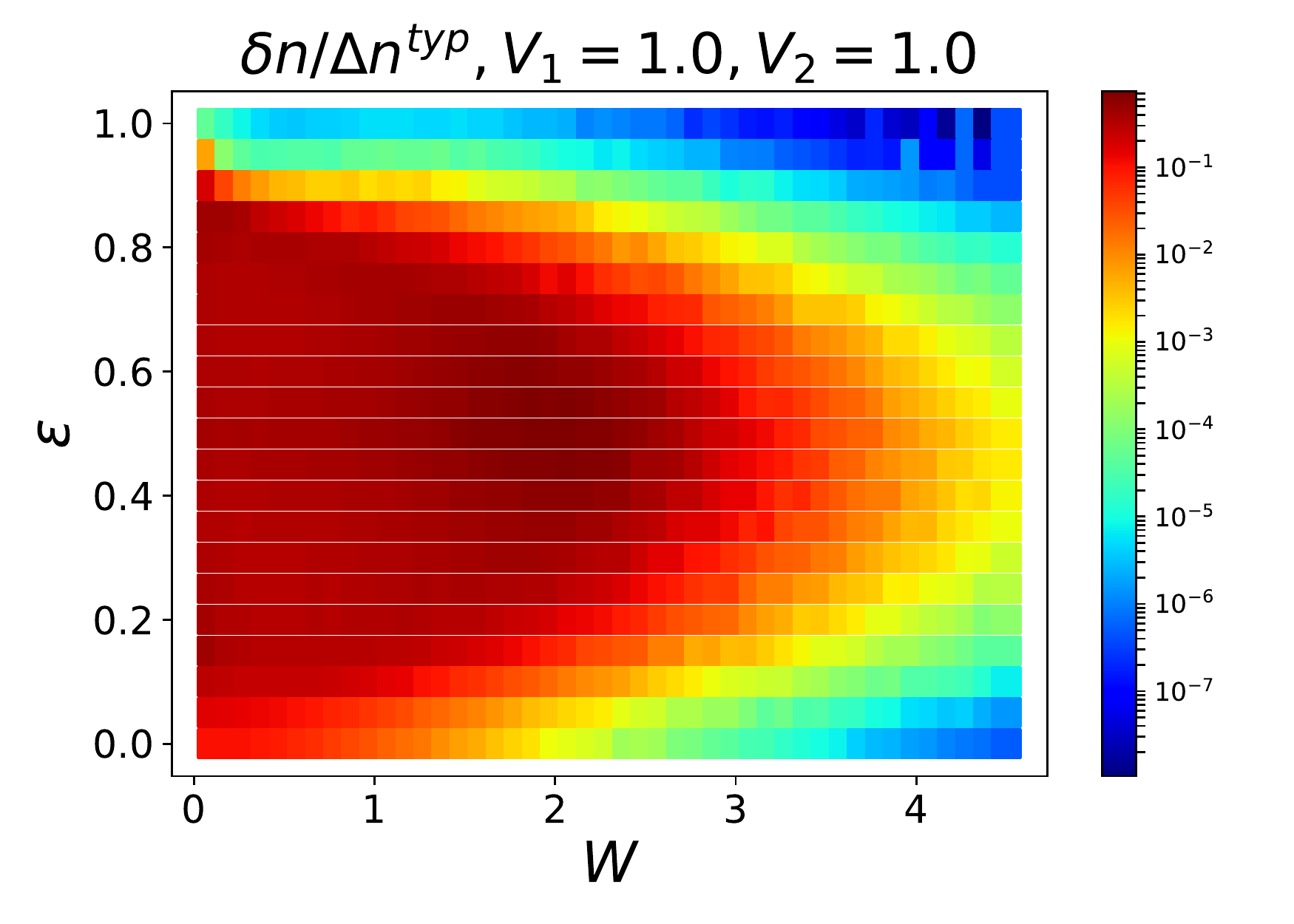}
  \end{subfigure}%
  \caption{(color online) typical average of
    $g_E=\frac{\delta E}{ \Delta E}$ (upper panel) and
    $g_n=\frac{\delta n}{ \Delta n}$ (lower panel) for the entire
    spectrum of the energy as we change the disorder strength $W$ for
    the case of having both NN and NNN interactions corresponding to
    $V_1=1, V_2=1$ in Eq. (\ref{H}). We set $L=14, N=7$. We take
    typical average over altogether $2000$ samples for each data
    point. \label{fig:shift_E_rho_nnn}}
\end{figure}

It is also instructive to apply our method of characterizing ETH-MBL
phase transition to the case of having both NN and NNN interactions,
corresponds to $V_1=1, V_2=1$ in Eq. (\ref{H}). Having NNN interaction
in addition to the NN interaction makes localization harder; i.e. we
expect that a larger amount of disorder is required to make the state
localized at each energy, and thus transition from ETH to MBL phase
happens at a larger value of $W_c$ compare to the NN case. Obtained
results of $g_E$ and $g_n$ for the case of $V_1=1, V_2=1$ are plotted
in Figs. \ref{fig:shift_n_nnn} and \ref{fig:shift_E_rho_nnn}. As we
expected, transition point moves to larger values of disorder
strength. We also see that the transition points become more
asymmetric compare to the NN case. Moreover, there is no phase
transition between ETH and MBL for states with the largest
$\epsilon$'s, and those states are localized with a non-zero disorder
strength.

\section{Conclusion and outlook}\label{conclusion}
Regarding the ETH-MBL phase transition in random interacting systems,
finding the phase characterizations are one of main research today. In
this work, we introduced new methods for characterizing the phase
transition; namely, we studied the response of the system to the
boundary conditions. For free fermions, the effect of change in
boundary conditions on single-particle
eigen-energies\cite{Edwards_1972}, as well as on the single-particle
density matrix\cite{PhysRevB.96.045123} has been studied
before. Extended eigenstates feel what happens at the boundary, while
changes in boundary conditions are not reflected in the localized
phase. We extend this idea to the interacting case. In particular, we
changed the boundary conditions between periodic and antiperiodic;
then, we studied the echo of these changes in the energy of the system
and eigenvalues of the single-particle density matrix. We used these
characterizations for the case of a 1D model with nearest-neighbor
interaction, nearest-neighbor hopping, and disorder which is added by
random onsite energies. This model has been studied before, and we
know approximately the phase transition point. We could identify the
ETH phase with significant shifts in the system's energy and
significant shifts in the occupation numbers (both shifts are on the
order of corresponding level spacing). In contrast, the MBL phase has
a vanishing response to the change in the boundary
conditions. Furthermore, we added extra next-nearest-neighbor
interactions and studied its effects on the ETH-MBL phase transition.

\acknowledgements

This work was supported by the Iran National Science Foundation (INSF)
and the University of Mazandaran (M. P.).

\bibliographystyle{apsrev4-1}
\bibliography{/home/cms/Dropbox/physics/Bib/reference.bib}

\begin{thebibliography}{62}%
\makeatletter
\providecommand \@ifxundefined [1]{%
 \@ifx{#1\undefined}
}%
\providecommand \@ifnum [1]{%
 \ifnum #1\expandafter \@firstoftwo
 \else \expandafter \@secondoftwo
 \fi
}%
\providecommand \@ifx [1]{%
 \ifx #1\expandafter \@firstoftwo
 \else \expandafter \@secondoftwo
 \fi
}%
\providecommand \natexlab [1]{#1}%
\providecommand \enquote  [1]{``#1''}%
\providecommand \bibnamefont  [1]{#1}%
\providecommand \bibfnamefont [1]{#1}%
\providecommand \citenamefont [1]{#1}%
\providecommand \href@noop [0]{\@secondoftwo}%
\providecommand \href [0]{\begingroup \@sanitize@url \@href}%
\providecommand \@href[1]{\@@startlink{#1}\@@href}%
\providecommand \@@href[1]{\endgroup#1\@@endlink}%
\providecommand \@sanitize@url [0]{\catcode `\\12\catcode `\$12\catcode
  `\&12\catcode `\#12\catcode `\^12\catcode `\_12\catcode `\%12\relax}%
\providecommand \@@startlink[1]{}%
\providecommand \@@endlink[0]{}%
\providecommand \url  [0]{\begingroup\@sanitize@url \@url }%
\providecommand \@url [1]{\endgroup\@href {#1}{\urlprefix }}%
\providecommand \urlprefix  [0]{URL }%
\providecommand \Eprint [0]{\href }%
\providecommand \doibase [0]{http://dx.doi.org/}%
\providecommand \selectlanguage [0]{\@gobble}%
\providecommand \bibinfo  [0]{\@secondoftwo}%
\providecommand \bibfield  [0]{\@secondoftwo}%
\providecommand \translation [1]{[#1]}%
\providecommand \BibitemOpen [0]{}%
\providecommand \bibitemStop [0]{}%
\providecommand \bibitemNoStop [0]{.\EOS\space}%
\providecommand \EOS [0]{\spacefactor3000\relax}%
\providecommand \BibitemShut  [1]{\csname bibitem#1\endcsname}%
\let\auto@bib@innerbib\@empty
\bibitem [{\citenamefont {Anderson}(1958)}]{PhysRev.109.1492}%
  \BibitemOpen
  \bibfield  {author} {\bibinfo {author} {\bibfnamefont {P.~W.}\ \bibnamefont
  {Anderson}},\ }\href {\doibase 10.1103/PhysRev.109.1492} {\bibfield
  {journal} {\bibinfo  {journal} {Phys. Rev.}\ }\textbf {\bibinfo {volume}
  {109}},\ \bibinfo {pages} {1492} (\bibinfo {year} {1958})}\BibitemShut
  {NoStop}%
\bibitem [{\citenamefont {Horodecki}\ \emph {et~al.}(2009)\citenamefont
  {Horodecki}, \citenamefont {Horodecki}, \citenamefont {Horodecki},\ and\
  \citenamefont {Horodecki}}]{RevModPhys.81.865}%
  \BibitemOpen
  \bibfield  {author} {\bibinfo {author} {\bibfnamefont {R.}~\bibnamefont
  {Horodecki}}, \bibinfo {author} {\bibfnamefont {P.}~\bibnamefont
  {Horodecki}}, \bibinfo {author} {\bibfnamefont {M.}~\bibnamefont
  {Horodecki}}, \ and\ \bibinfo {author} {\bibfnamefont {K.}~\bibnamefont
  {Horodecki}},\ }\href {\doibase 10.1103/RevModPhys.81.865} {\bibfield
  {journal} {\bibinfo  {journal} {Rev. Mod. Phys.}\ }\textbf {\bibinfo {volume}
  {81}},\ \bibinfo {pages} {865} (\bibinfo {year} {2009})}\BibitemShut
  {NoStop}%
\bibitem [{\citenamefont {Evers}\ and\ \citenamefont
  {Mirlin}(2008)}]{RevModPhys.80.1355}%
  \BibitemOpen
  \bibfield  {author} {\bibinfo {author} {\bibfnamefont {F.}~\bibnamefont
  {Evers}}\ and\ \bibinfo {author} {\bibfnamefont {A.~D.}\ \bibnamefont
  {Mirlin}},\ }\href {\doibase 10.1103/RevModPhys.80.1355} {\bibfield
  {journal} {\bibinfo  {journal} {Rev. Mod. Phys.}\ }\textbf {\bibinfo {volume}
  {80}},\ \bibinfo {pages} {1355} (\bibinfo {year} {2008})}\BibitemShut
  {NoStop}%
\bibitem [{\citenamefont {MacKinnon}\ and\ \citenamefont
  {Kramer}(1981)}]{PhysRevLett.47.1546}%
  \BibitemOpen
  \bibfield  {author} {\bibinfo {author} {\bibfnamefont {A.}~\bibnamefont
  {MacKinnon}}\ and\ \bibinfo {author} {\bibfnamefont {B.}~\bibnamefont
  {Kramer}},\ }\href {\doibase 10.1103/PhysRevLett.47.1546} {\bibfield
  {journal} {\bibinfo  {journal} {Phys. Rev. Lett.}\ }\textbf {\bibinfo
  {volume} {47}},\ \bibinfo {pages} {1546} (\bibinfo {year}
  {1981})}\BibitemShut {NoStop}%
\bibitem [{\citenamefont {Economou}\ and\ \citenamefont
  {Cohen}(1972)}]{PhysRevB.5.2931}%
  \BibitemOpen
  \bibfield  {author} {\bibinfo {author} {\bibfnamefont {E.~N.}\ \bibnamefont
  {Economou}}\ and\ \bibinfo {author} {\bibfnamefont {M.~H.}\ \bibnamefont
  {Cohen}},\ }\href {\doibase 10.1103/PhysRevB.5.2931} {\bibfield  {journal}
  {\bibinfo  {journal} {Phys. Rev. B}\ }\textbf {\bibinfo {volume} {5}},\
  \bibinfo {pages} {2931} (\bibinfo {year} {1972})}\BibitemShut {NoStop}%
\bibitem [{\citenamefont {Izrailev}\ and\ \citenamefont
  {Krokhin}(1999)}]{PhysRevLett.82.4062}%
  \BibitemOpen
  \bibfield  {author} {\bibinfo {author} {\bibfnamefont {F.~M.}\ \bibnamefont
  {Izrailev}}\ and\ \bibinfo {author} {\bibfnamefont {A.~A.}\ \bibnamefont
  {Krokhin}},\ }\href {\doibase 10.1103/PhysRevLett.82.4062} {\bibfield
  {journal} {\bibinfo  {journal} {Phys. Rev. Lett.}\ }\textbf {\bibinfo
  {volume} {82}},\ \bibinfo {pages} {4062} (\bibinfo {year}
  {1999})}\BibitemShut {NoStop}%
\bibitem [{\citenamefont {Nandkishore}\ and\ \citenamefont
  {Huse}(2015)}]{doi:10.1146/annurev-conmatphys-031214-014726}%
  \BibitemOpen
  \bibfield  {author} {\bibinfo {author} {\bibfnamefont {R.}~\bibnamefont
  {Nandkishore}}\ and\ \bibinfo {author} {\bibfnamefont {D.~A.}\ \bibnamefont
  {Huse}},\ }\href {\doibase 10.1146/annurev-conmatphys-031214-014726}
  {\bibfield  {journal} {\bibinfo  {journal} {Annual Review of Condensed Matter
  Physics}\ }\textbf {\bibinfo {volume} {6}},\ \bibinfo {pages} {15} (\bibinfo
  {year} {2015})},\ \Eprint
  {http://arxiv.org/abs/https://doi.org/10.1146/annurev-conmatphys-031214-014726}
  {https://doi.org/10.1146/annurev-conmatphys-031214-014726} \BibitemShut
  {NoStop}%
\bibitem [{\citenamefont {Altman}(2018)}]{Altman2018}%
  \BibitemOpen
  \bibfield  {author} {\bibinfo {author} {\bibfnamefont {E.}~\bibnamefont
  {Altman}},\ }\href {\doibase 10.1038/s41567-018-0305-7} {\bibfield  {journal}
  {\bibinfo  {journal} {Nature Physics}\ }\textbf {\bibinfo {volume} {14}},\
  \bibinfo {pages} {979} (\bibinfo {year} {2018})}\BibitemShut {NoStop}%
\bibitem [{\citenamefont {Abanin}\ \emph {et~al.}(2019)\citenamefont {Abanin},
  \citenamefont {Altman}, \citenamefont {Bloch},\ and\ \citenamefont
  {Serbyn}}]{RevModPhys.91.021001}%
  \BibitemOpen
  \bibfield  {author} {\bibinfo {author} {\bibfnamefont {D.~A.}\ \bibnamefont
  {Abanin}}, \bibinfo {author} {\bibfnamefont {E.}~\bibnamefont {Altman}},
  \bibinfo {author} {\bibfnamefont {I.}~\bibnamefont {Bloch}}, \ and\ \bibinfo
  {author} {\bibfnamefont {M.}~\bibnamefont {Serbyn}},\ }\href {\doibase
  10.1103/RevModPhys.91.021001} {\bibfield  {journal} {\bibinfo  {journal}
  {Rev. Mod. Phys.}\ }\textbf {\bibinfo {volume} {91}},\ \bibinfo {pages}
  {021001} (\bibinfo {year} {2019})}\BibitemShut {NoStop}%
\bibitem [{\citenamefont {Basko}\ \emph {et~al.}(2006)\citenamefont {Basko},
  \citenamefont {Aleiner},\ and\ \citenamefont {Altshuler}}]{BASKO20061126}%
  \BibitemOpen
  \bibfield  {author} {\bibinfo {author} {\bibfnamefont {D.}~\bibnamefont
  {Basko}}, \bibinfo {author} {\bibfnamefont {I.}~\bibnamefont {Aleiner}}, \
  and\ \bibinfo {author} {\bibfnamefont {B.}~\bibnamefont {Altshuler}},\ }\href
  {\doibase https://doi.org/10.1016/j.aop.2005.11.014} {\bibfield  {journal}
  {\bibinfo  {journal} {Annals of Physics}\ }\textbf {\bibinfo {volume}
  {321}},\ \bibinfo {pages} {1126 } (\bibinfo {year} {2006})}\BibitemShut
  {NoStop}%
\bibitem [{\citenamefont {Parameswaran}\ \emph {et~al.}(2017)\citenamefont
  {Parameswaran}, \citenamefont {Potter},\ and\ \citenamefont
  {Vasseur}}]{doi:10.1002/andp.201600302}%
  \BibitemOpen
  \bibfield  {author} {\bibinfo {author} {\bibfnamefont {S.~A.}\ \bibnamefont
  {Parameswaran}}, \bibinfo {author} {\bibfnamefont {A.~C.}\ \bibnamefont
  {Potter}}, \ and\ \bibinfo {author} {\bibfnamefont {R.}~\bibnamefont
  {Vasseur}},\ }\href {\doibase 10.1002/andp.201600302} {\bibfield  {journal}
  {\bibinfo  {journal} {Annalen der Physik}\ }\textbf {\bibinfo {volume}
  {529}},\ \bibinfo {pages} {1600302} (\bibinfo {year} {2017})},\ \Eprint
  {http://arxiv.org/abs/https://onlinelibrary.wiley.com/doi/pdf/10.1002/andp.201600302}
  {https://onlinelibrary.wiley.com/doi/pdf/10.1002/andp.201600302} \BibitemShut
  {NoStop}%
\bibitem [{\citenamefont {Alet}\ and\ \citenamefont
  {Laflorencie}(2018)}]{ALET2018498}%
  \BibitemOpen
  \bibfield  {author} {\bibinfo {author} {\bibfnamefont {F.}~\bibnamefont
  {Alet}}\ and\ \bibinfo {author} {\bibfnamefont {N.}~\bibnamefont
  {Laflorencie}},\ }\href {\doibase https://doi.org/10.1016/j.crhy.2018.03.003}
  {\bibfield  {journal} {\bibinfo  {journal} {Comptes Rendus Physique}\
  }\textbf {\bibinfo {volume} {19}},\ \bibinfo {pages} {498 } (\bibinfo {year}
  {2018})},\ \bibinfo {note} {quantum simulation / Simulation
  quantique}\BibitemShut {NoStop}%
\bibitem [{\citenamefont {Altman}\ and\ \citenamefont
  {Vosk}(2015)}]{doi:10.1146/annurev-conmatphys-031214-014701}%
  \BibitemOpen
  \bibfield  {author} {\bibinfo {author} {\bibfnamefont {E.}~\bibnamefont
  {Altman}}\ and\ \bibinfo {author} {\bibfnamefont {R.}~\bibnamefont {Vosk}},\
  }\href {\doibase 10.1146/annurev-conmatphys-031214-014701} {\bibfield
  {journal} {\bibinfo  {journal} {Annual Review of Condensed Matter Physics}\
  }\textbf {\bibinfo {volume} {6}},\ \bibinfo {pages} {383} (\bibinfo {year}
  {2015})},\ \Eprint
  {http://arxiv.org/abs/https://doi.org/10.1146/annurev-conmatphys-031214-014701}
  {https://doi.org/10.1146/annurev-conmatphys-031214-014701} \BibitemShut
  {NoStop}%
\bibitem [{\citenamefont {Deutsch}(1991)}]{PhysRevA.43.2046}%
  \BibitemOpen
  \bibfield  {author} {\bibinfo {author} {\bibfnamefont {J.~M.}\ \bibnamefont
  {Deutsch}},\ }\href {\doibase 10.1103/PhysRevA.43.2046} {\bibfield  {journal}
  {\bibinfo  {journal} {Phys. Rev. A}\ }\textbf {\bibinfo {volume} {43}},\
  \bibinfo {pages} {2046} (\bibinfo {year} {1991})}\BibitemShut {NoStop}%
\bibitem [{\citenamefont {Srednicki}(1994)}]{PhysRevE.50.888}%
  \BibitemOpen
  \bibfield  {author} {\bibinfo {author} {\bibfnamefont {M.}~\bibnamefont
  {Srednicki}},\ }\href {\doibase 10.1103/PhysRevE.50.888} {\bibfield
  {journal} {\bibinfo  {journal} {Phys. Rev. E}\ }\textbf {\bibinfo {volume}
  {50}},\ \bibinfo {pages} {888} (\bibinfo {year} {1994})}\BibitemShut
  {NoStop}%
\bibitem [{\citenamefont {Rigol}\ \emph {et~al.}(2008)\citenamefont {Rigol},
  \citenamefont {Dunjko},\ and\ \citenamefont {Olshanii}}]{Rigol2008}%
  \BibitemOpen
  \bibfield  {author} {\bibinfo {author} {\bibfnamefont {M.}~\bibnamefont
  {Rigol}}, \bibinfo {author} {\bibfnamefont {V.}~\bibnamefont {Dunjko}}, \
  and\ \bibinfo {author} {\bibfnamefont {M.}~\bibnamefont {Olshanii}},\ }\href
  {https://doi.org/10.1038/nature06838} {\bibfield  {journal} {\bibinfo
  {journal} {Nature}\ }\textbf {\bibinfo {volume} {452}},\ \bibinfo {pages}
  {854 EP } (\bibinfo {year} {2008})}\BibitemShut {NoStop}%
\bibitem [{\citenamefont {Deutsch}(2018)}]{Deutsch_2018}%
  \BibitemOpen
  \bibfield  {author} {\bibinfo {author} {\bibfnamefont {J.~M.}\ \bibnamefont
  {Deutsch}},\ }\href {\doibase 10.1088/1361-6633/aac9f1} {\bibfield  {journal}
  {\bibinfo  {journal} {Reports on Progress in Physics}\ }\textbf {\bibinfo
  {volume} {81}},\ \bibinfo {pages} {082001} (\bibinfo {year}
  {2018})}\BibitemShut {NoStop}%
\bibitem [{\citenamefont {Baygan}\ \emph {et~al.}(2015)\citenamefont {Baygan},
  \citenamefont {Lim},\ and\ \citenamefont {Sheng}}]{PhysRevB.92.195153}%
  \BibitemOpen
  \bibfield  {author} {\bibinfo {author} {\bibfnamefont {E.}~\bibnamefont
  {Baygan}}, \bibinfo {author} {\bibfnamefont {S.~P.}\ \bibnamefont {Lim}}, \
  and\ \bibinfo {author} {\bibfnamefont {D.~N.}\ \bibnamefont {Sheng}},\ }\href
  {\doibase 10.1103/PhysRevB.92.195153} {\bibfield  {journal} {\bibinfo
  {journal} {Phys. Rev. B}\ }\textbf {\bibinfo {volume} {92}},\ \bibinfo
  {pages} {195153} (\bibinfo {year} {2015})}\BibitemShut {NoStop}%
\bibitem [{\citenamefont {Li}\ \emph {et~al.}(2015)\citenamefont {Li},
  \citenamefont {Ganeshan}, \citenamefont {Pixley},\ and\ \citenamefont
  {Das~Sarma}}]{PhysRevLett.115.186601}%
  \BibitemOpen
  \bibfield  {author} {\bibinfo {author} {\bibfnamefont {X.}~\bibnamefont
  {Li}}, \bibinfo {author} {\bibfnamefont {S.}~\bibnamefont {Ganeshan}},
  \bibinfo {author} {\bibfnamefont {J.~H.}\ \bibnamefont {Pixley}}, \ and\
  \bibinfo {author} {\bibfnamefont {S.}~\bibnamefont {Das~Sarma}},\ }\href
  {\doibase 10.1103/PhysRevLett.115.186601} {\bibfield  {journal} {\bibinfo
  {journal} {Phys. Rev. Lett.}\ }\textbf {\bibinfo {volume} {115}},\ \bibinfo
  {pages} {186601} (\bibinfo {year} {2015})}\BibitemShut {NoStop}%
\bibitem [{\citenamefont {Bloch}\ \emph {et~al.}(2008)\citenamefont {Bloch},
  \citenamefont {Dalibard},\ and\ \citenamefont {Zwerger}}]{RevModPhys.80.885}%
  \BibitemOpen
  \bibfield  {author} {\bibinfo {author} {\bibfnamefont {I.}~\bibnamefont
  {Bloch}}, \bibinfo {author} {\bibfnamefont {J.}~\bibnamefont {Dalibard}}, \
  and\ \bibinfo {author} {\bibfnamefont {W.}~\bibnamefont {Zwerger}},\ }\href
  {\doibase 10.1103/RevModPhys.80.885} {\bibfield  {journal} {\bibinfo
  {journal} {Rev. Mod. Phys.}\ }\textbf {\bibinfo {volume} {80}},\ \bibinfo
  {pages} {885} (\bibinfo {year} {2008})}\BibitemShut {NoStop}%
\bibitem [{\citenamefont {Smith}\ \emph {et~al.}(2016)\citenamefont {Smith},
  \citenamefont {Lee}, \citenamefont {Richerme}, \citenamefont {Neyenhuis},
  \citenamefont {Hess}, \citenamefont {Hauke}, \citenamefont {Heyl},
  \citenamefont {Huse},\ and\ \citenamefont {Monroe}}]{Smith2016}%
  \BibitemOpen
  \bibfield  {author} {\bibinfo {author} {\bibfnamefont {J.}~\bibnamefont
  {Smith}}, \bibinfo {author} {\bibfnamefont {A.}~\bibnamefont {Lee}}, \bibinfo
  {author} {\bibfnamefont {P.}~\bibnamefont {Richerme}}, \bibinfo {author}
  {\bibfnamefont {B.}~\bibnamefont {Neyenhuis}}, \bibinfo {author}
  {\bibfnamefont {P.~W.}\ \bibnamefont {Hess}}, \bibinfo {author}
  {\bibfnamefont {P.}~\bibnamefont {Hauke}}, \bibinfo {author} {\bibfnamefont
  {M.}~\bibnamefont {Heyl}}, \bibinfo {author} {\bibfnamefont {D.~A.}\
  \bibnamefont {Huse}}, \ and\ \bibinfo {author} {\bibfnamefont
  {C.}~\bibnamefont {Monroe}},\ }\href {https://doi.org/10.1038/nphys3783}
  {\bibfield  {journal} {\bibinfo  {journal} {Nature Physics}\ }\textbf
  {\bibinfo {volume} {12}},\ \bibinfo {pages} {907 EP } (\bibinfo {year}
  {2016})}\BibitemShut {NoStop}%
\bibitem [{\citenamefont {Langen}\ \emph {et~al.}(2015)\citenamefont {Langen},
  \citenamefont {Geiger},\ and\ \citenamefont
  {Schmiedmayer}}]{doi:10.1146/annurev-conmatphys-031214-014548}%
  \BibitemOpen
  \bibfield  {author} {\bibinfo {author} {\bibfnamefont {T.}~\bibnamefont
  {Langen}}, \bibinfo {author} {\bibfnamefont {R.}~\bibnamefont {Geiger}}, \
  and\ \bibinfo {author} {\bibfnamefont {J.}~\bibnamefont {Schmiedmayer}},\
  }\href {\doibase 10.1146/annurev-conmatphys-031214-014548} {\bibfield
  {journal} {\bibinfo  {journal} {Annual Review of Condensed Matter Physics}\
  }\textbf {\bibinfo {volume} {6}},\ \bibinfo {pages} {201} (\bibinfo {year}
  {2015})},\ \Eprint
  {http://arxiv.org/abs/https://doi.org/10.1146/annurev-conmatphys-031214-014548}
  {https://doi.org/10.1146/annurev-conmatphys-031214-014548} \BibitemShut
  {NoStop}%
\bibitem [{\citenamefont {Chin}\ \emph {et~al.}(2010)\citenamefont {Chin},
  \citenamefont {Grimm}, \citenamefont {Julienne},\ and\ \citenamefont
  {Tiesinga}}]{RevModPhys.82.1225}%
  \BibitemOpen
  \bibfield  {author} {\bibinfo {author} {\bibfnamefont {C.}~\bibnamefont
  {Chin}}, \bibinfo {author} {\bibfnamefont {R.}~\bibnamefont {Grimm}},
  \bibinfo {author} {\bibfnamefont {P.}~\bibnamefont {Julienne}}, \ and\
  \bibinfo {author} {\bibfnamefont {E.}~\bibnamefont {Tiesinga}},\ }\href
  {\doibase 10.1103/RevModPhys.82.1225} {\bibfield  {journal} {\bibinfo
  {journal} {Rev. Mod. Phys.}\ }\textbf {\bibinfo {volume} {82}},\ \bibinfo
  {pages} {1225} (\bibinfo {year} {2010})}\BibitemShut {NoStop}%
\bibitem [{\citenamefont {Kaufman}\ \emph {et~al.}(2016)\citenamefont
  {Kaufman}, \citenamefont {Tai}, \citenamefont {Lukin}, \citenamefont
  {Rispoli}, \citenamefont {Schittko}, \citenamefont {Preiss},\ and\
  \citenamefont {Greiner}}]{Kaufman794}%
  \BibitemOpen
  \bibfield  {author} {\bibinfo {author} {\bibfnamefont {A.~M.}\ \bibnamefont
  {Kaufman}}, \bibinfo {author} {\bibfnamefont {M.~E.}\ \bibnamefont {Tai}},
  \bibinfo {author} {\bibfnamefont {A.}~\bibnamefont {Lukin}}, \bibinfo
  {author} {\bibfnamefont {M.}~\bibnamefont {Rispoli}}, \bibinfo {author}
  {\bibfnamefont {R.}~\bibnamefont {Schittko}}, \bibinfo {author}
  {\bibfnamefont {P.~M.}\ \bibnamefont {Preiss}}, \ and\ \bibinfo {author}
  {\bibfnamefont {M.}~\bibnamefont {Greiner}},\ }\href {\doibase
  10.1126/science.aaf6725} {\bibfield  {journal} {\bibinfo  {journal}
  {Science}\ }\textbf {\bibinfo {volume} {353}},\ \bibinfo {pages} {794}
  (\bibinfo {year} {2016})},\ \Eprint
  {http://arxiv.org/abs/http://science.sciencemag.org/content/353/6301/794.full.pdf}
  {http://science.sciencemag.org/content/353/6301/794.full.pdf} \BibitemShut
  {NoStop}%
\bibitem [{\citenamefont {Bordia}\ \emph {et~al.}(2017)\citenamefont {Bordia},
  \citenamefont {L{\"u}schen}, \citenamefont {Schneider}, \citenamefont
  {Knap},\ and\ \citenamefont {Bloch}}]{Bordia2017}%
  \BibitemOpen
  \bibfield  {author} {\bibinfo {author} {\bibfnamefont {P.}~\bibnamefont
  {Bordia}}, \bibinfo {author} {\bibfnamefont {H.}~\bibnamefont {L{\"u}schen}},
  \bibinfo {author} {\bibfnamefont {U.}~\bibnamefont {Schneider}}, \bibinfo
  {author} {\bibfnamefont {M.}~\bibnamefont {Knap}}, \ and\ \bibinfo {author}
  {\bibfnamefont {I.}~\bibnamefont {Bloch}},\ }\href
  {https://doi.org/10.1038/nphys4020} {\bibfield  {journal} {\bibinfo
  {journal} {Nature Physics}\ }\textbf {\bibinfo {volume} {13}},\ \bibinfo
  {pages} {460 EP } (\bibinfo {year} {2017})},\ \bibinfo {note}
  {article}\BibitemShut {NoStop}%
\bibitem [{\citenamefont {Choi}\ \emph {et~al.}(2016)\citenamefont {Choi},
  \citenamefont {Hild}, \citenamefont {Zeiher}, \citenamefont {Schau{\ss}},
  \citenamefont {Rubio-Abadal}, \citenamefont {Yefsah}, \citenamefont
  {Khemani}, \citenamefont {Huse}, \citenamefont {Bloch},\ and\ \citenamefont
  {Gross}}]{Choi1547}%
  \BibitemOpen
  \bibfield  {author} {\bibinfo {author} {\bibfnamefont {J.-y.}\ \bibnamefont
  {Choi}}, \bibinfo {author} {\bibfnamefont {S.}~\bibnamefont {Hild}}, \bibinfo
  {author} {\bibfnamefont {J.}~\bibnamefont {Zeiher}}, \bibinfo {author}
  {\bibfnamefont {P.}~\bibnamefont {Schau{\ss}}}, \bibinfo {author}
  {\bibfnamefont {A.}~\bibnamefont {Rubio-Abadal}}, \bibinfo {author}
  {\bibfnamefont {T.}~\bibnamefont {Yefsah}}, \bibinfo {author} {\bibfnamefont
  {V.}~\bibnamefont {Khemani}}, \bibinfo {author} {\bibfnamefont {D.~A.}\
  \bibnamefont {Huse}}, \bibinfo {author} {\bibfnamefont {I.}~\bibnamefont
  {Bloch}}, \ and\ \bibinfo {author} {\bibfnamefont {C.}~\bibnamefont
  {Gross}},\ }\href {\doibase 10.1126/science.aaf8834} {\bibfield  {journal}
  {\bibinfo  {journal} {Science}\ }\textbf {\bibinfo {volume} {352}},\ \bibinfo
  {pages} {1547} (\bibinfo {year} {2016})},\ \Eprint
  {http://arxiv.org/abs/http://science.sciencemag.org/content/352/6293/1547.full.pdf}
  {http://science.sciencemag.org/content/352/6293/1547.full.pdf} \BibitemShut
  {NoStop}%
\bibitem [{\citenamefont {Bordia}\ \emph {et~al.}(2016)\citenamefont {Bordia},
  \citenamefont {L\"uschen}, \citenamefont {Hodgman}, \citenamefont
  {Schreiber}, \citenamefont {Bloch},\ and\ \citenamefont
  {Schneider}}]{PhysRevLett.116.140401}%
  \BibitemOpen
  \bibfield  {author} {\bibinfo {author} {\bibfnamefont {P.}~\bibnamefont
  {Bordia}}, \bibinfo {author} {\bibfnamefont {H.~P.}\ \bibnamefont
  {L\"uschen}}, \bibinfo {author} {\bibfnamefont {S.~S.}\ \bibnamefont
  {Hodgman}}, \bibinfo {author} {\bibfnamefont {M.}~\bibnamefont {Schreiber}},
  \bibinfo {author} {\bibfnamefont {I.}~\bibnamefont {Bloch}}, \ and\ \bibinfo
  {author} {\bibfnamefont {U.}~\bibnamefont {Schneider}},\ }\href {\doibase
  10.1103/PhysRevLett.116.140401} {\bibfield  {journal} {\bibinfo  {journal}
  {Phys. Rev. Lett.}\ }\textbf {\bibinfo {volume} {116}},\ \bibinfo {pages}
  {140401} (\bibinfo {year} {2016})}\BibitemShut {NoStop}%
\bibitem [{\citenamefont {Schreiber}\ \emph {et~al.}(2015)\citenamefont
  {Schreiber}, \citenamefont {Hodgman}, \citenamefont {Bordia}, \citenamefont
  {L{\"u}schen}, \citenamefont {Fischer}, \citenamefont {Vosk}, \citenamefont
  {Altman}, \citenamefont {Schneider},\ and\ \citenamefont
  {Bloch}}]{Schreiber842}%
  \BibitemOpen
  \bibfield  {author} {\bibinfo {author} {\bibfnamefont {M.}~\bibnamefont
  {Schreiber}}, \bibinfo {author} {\bibfnamefont {S.~S.}\ \bibnamefont
  {Hodgman}}, \bibinfo {author} {\bibfnamefont {P.}~\bibnamefont {Bordia}},
  \bibinfo {author} {\bibfnamefont {H.~P.}\ \bibnamefont {L{\"u}schen}},
  \bibinfo {author} {\bibfnamefont {M.~H.}\ \bibnamefont {Fischer}}, \bibinfo
  {author} {\bibfnamefont {R.}~\bibnamefont {Vosk}}, \bibinfo {author}
  {\bibfnamefont {E.}~\bibnamefont {Altman}}, \bibinfo {author} {\bibfnamefont
  {U.}~\bibnamefont {Schneider}}, \ and\ \bibinfo {author} {\bibfnamefont
  {I.}~\bibnamefont {Bloch}},\ }\href {\doibase 10.1126/science.aaa7432}
  {\bibfield  {journal} {\bibinfo  {journal} {Science}\ }\textbf {\bibinfo
  {volume} {349}},\ \bibinfo {pages} {842} (\bibinfo {year} {2015})},\ \Eprint
  {http://arxiv.org/abs/http://science.sciencemag.org/content/349/6250/842.full.pdf}
  {http://science.sciencemag.org/content/349/6250/842.full.pdf} \BibitemShut
  {NoStop}%
\bibitem [{\citenamefont {Sapienza}\ \emph {et~al.}(2010)\citenamefont
  {Sapienza}, \citenamefont {Thyrrestrup}, \citenamefont {Stobbe},
  \citenamefont {Garcia}, \citenamefont {Smolka},\ and\ \citenamefont
  {Lodahl}}]{Sapienza1352}%
  \BibitemOpen
  \bibfield  {author} {\bibinfo {author} {\bibfnamefont {L.}~\bibnamefont
  {Sapienza}}, \bibinfo {author} {\bibfnamefont {H.}~\bibnamefont
  {Thyrrestrup}}, \bibinfo {author} {\bibfnamefont {S.}~\bibnamefont {Stobbe}},
  \bibinfo {author} {\bibfnamefont {P.~D.}\ \bibnamefont {Garcia}}, \bibinfo
  {author} {\bibfnamefont {S.}~\bibnamefont {Smolka}}, \ and\ \bibinfo {author}
  {\bibfnamefont {P.}~\bibnamefont {Lodahl}},\ }\href {\doibase
  10.1126/science.1185080} {\bibfield  {journal} {\bibinfo  {journal}
  {Science}\ }\textbf {\bibinfo {volume} {327}},\ \bibinfo {pages} {1352}
  (\bibinfo {year} {2010})},\ \Eprint
  {http://arxiv.org/abs/http://science.sciencemag.org/content/327/5971/1352.full.pdf}
  {http://science.sciencemag.org/content/327/5971/1352.full.pdf} \BibitemShut
  {NoStop}%
\bibitem [{\citenamefont {Papić}\ \emph {et~al.}(2015)\citenamefont {Papić},
  \citenamefont {Stoudenmire},\ and\ \citenamefont {Abanin}}]{PAPIC2015714}%
  \BibitemOpen
  \bibfield  {author} {\bibinfo {author} {\bibfnamefont {Z.}~\bibnamefont
  {Papić}}, \bibinfo {author} {\bibfnamefont {E.~M.}\ \bibnamefont
  {Stoudenmire}}, \ and\ \bibinfo {author} {\bibfnamefont {D.~A.}\ \bibnamefont
  {Abanin}},\ }\href {\doibase https://doi.org/10.1016/j.aop.2015.08.024}
  {\bibfield  {journal} {\bibinfo  {journal} {Annals of Physics}\ }\textbf
  {\bibinfo {volume} {362}},\ \bibinfo {pages} {714 } (\bibinfo {year}
  {2015})}\BibitemShut {NoStop}%
\bibitem [{\citenamefont {Li}\ \emph {et~al.}(2017)\citenamefont {Li},
  \citenamefont {Deng}, \citenamefont {Wu},\ and\ \citenamefont
  {Das~Sarma}}]{PhysRevB.95.020201}%
  \BibitemOpen
  \bibfield  {author} {\bibinfo {author} {\bibfnamefont {X.}~\bibnamefont
  {Li}}, \bibinfo {author} {\bibfnamefont {D.-L.}\ \bibnamefont {Deng}},
  \bibinfo {author} {\bibfnamefont {Y.-L.}\ \bibnamefont {Wu}}, \ and\ \bibinfo
  {author} {\bibfnamefont {S.}~\bibnamefont {Das~Sarma}},\ }\href {\doibase
  10.1103/PhysRevB.95.020201} {\bibfield  {journal} {\bibinfo  {journal} {Phys.
  Rev. B}\ }\textbf {\bibinfo {volume} {95}},\ \bibinfo {pages} {020201}
  (\bibinfo {year} {2017})}\BibitemShut {NoStop}%
\bibitem [{\citenamefont {Bar~Lev}\ \emph {et~al.}(2016)\citenamefont
  {Bar~Lev}, \citenamefont {Reichman},\ and\ \citenamefont
  {Sagi}}]{PhysRevB.94.201116}%
  \BibitemOpen
  \bibfield  {author} {\bibinfo {author} {\bibfnamefont {Y.}~\bibnamefont
  {Bar~Lev}}, \bibinfo {author} {\bibfnamefont {D.~R.}\ \bibnamefont
  {Reichman}}, \ and\ \bibinfo {author} {\bibfnamefont {Y.}~\bibnamefont
  {Sagi}},\ }\href {\doibase 10.1103/PhysRevB.94.201116} {\bibfield  {journal}
  {\bibinfo  {journal} {Phys. Rev. B}\ }\textbf {\bibinfo {volume} {94}},\
  \bibinfo {pages} {201116} (\bibinfo {year} {2016})}\BibitemShut {NoStop}%
\bibitem [{\citenamefont {Yao}\ \emph {et~al.}(2016)\citenamefont {Yao},
  \citenamefont {Laumann}, \citenamefont {Cirac}, \citenamefont {Lukin},\ and\
  \citenamefont {Moore}}]{PhysRevLett.117.240601}%
  \BibitemOpen
  \bibfield  {author} {\bibinfo {author} {\bibfnamefont {N.~Y.}\ \bibnamefont
  {Yao}}, \bibinfo {author} {\bibfnamefont {C.~R.}\ \bibnamefont {Laumann}},
  \bibinfo {author} {\bibfnamefont {J.~I.}\ \bibnamefont {Cirac}}, \bibinfo
  {author} {\bibfnamefont {M.~D.}\ \bibnamefont {Lukin}}, \ and\ \bibinfo
  {author} {\bibfnamefont {J.~E.}\ \bibnamefont {Moore}},\ }\href {\doibase
  10.1103/PhysRevLett.117.240601} {\bibfield  {journal} {\bibinfo  {journal}
  {Phys. Rev. Lett.}\ }\textbf {\bibinfo {volume} {117}},\ \bibinfo {pages}
  {240601} (\bibinfo {year} {2016})}\BibitemShut {NoStop}%
\bibitem [{\citenamefont {Smith}\ \emph {et~al.}(2017)\citenamefont {Smith},
  \citenamefont {Knolle}, \citenamefont {Kovrizhin},\ and\ \citenamefont
  {Moessner}}]{PhysRevLett.118.266601}%
  \BibitemOpen
  \bibfield  {author} {\bibinfo {author} {\bibfnamefont {A.}~\bibnamefont
  {Smith}}, \bibinfo {author} {\bibfnamefont {J.}~\bibnamefont {Knolle}},
  \bibinfo {author} {\bibfnamefont {D.~L.}\ \bibnamefont {Kovrizhin}}, \ and\
  \bibinfo {author} {\bibfnamefont {R.}~\bibnamefont {Moessner}},\ }\href
  {\doibase 10.1103/PhysRevLett.118.266601} {\bibfield  {journal} {\bibinfo
  {journal} {Phys. Rev. Lett.}\ }\textbf {\bibinfo {volume} {118}},\ \bibinfo
  {pages} {266601} (\bibinfo {year} {2017})}\BibitemShut {NoStop}%
\bibitem [{\citenamefont {De~Roeck}\ and\ \citenamefont
  {Huveneers}(2014)}]{PhysRevB.90.165137}%
  \BibitemOpen
  \bibfield  {author} {\bibinfo {author} {\bibfnamefont {W.}~\bibnamefont
  {De~Roeck}}\ and\ \bibinfo {author} {\bibfnamefont {F.~m.~c.}\ \bibnamefont
  {Huveneers}},\ }\href {\doibase 10.1103/PhysRevB.90.165137} {\bibfield
  {journal} {\bibinfo  {journal} {Phys. Rev. B}\ }\textbf {\bibinfo {volume}
  {90}},\ \bibinfo {pages} {165137} (\bibinfo {year} {2014})}\BibitemShut
  {NoStop}%
\bibitem [{\citenamefont {Modak}\ and\ \citenamefont
  {Mukerjee}(2015)}]{PhysRevLett.115.230401}%
  \BibitemOpen
  \bibfield  {author} {\bibinfo {author} {\bibfnamefont {R.}~\bibnamefont
  {Modak}}\ and\ \bibinfo {author} {\bibfnamefont {S.}~\bibnamefont
  {Mukerjee}},\ }\href {\doibase 10.1103/PhysRevLett.115.230401} {\bibfield
  {journal} {\bibinfo  {journal} {Phys. Rev. Lett.}\ }\textbf {\bibinfo
  {volume} {115}},\ \bibinfo {pages} {230401} (\bibinfo {year}
  {2015})}\BibitemShut {NoStop}%
\bibitem [{\citenamefont {Li}\ \emph {et~al.}(2016)\citenamefont {Li},
  \citenamefont {Pixley}, \citenamefont {Deng}, \citenamefont {Ganeshan},\ and\
  \citenamefont {Das~Sarma}}]{PhysRevB.93.184204}%
  \BibitemOpen
  \bibfield  {author} {\bibinfo {author} {\bibfnamefont {X.}~\bibnamefont
  {Li}}, \bibinfo {author} {\bibfnamefont {J.~H.}\ \bibnamefont {Pixley}},
  \bibinfo {author} {\bibfnamefont {D.-L.}\ \bibnamefont {Deng}}, \bibinfo
  {author} {\bibfnamefont {S.}~\bibnamefont {Ganeshan}}, \ and\ \bibinfo
  {author} {\bibfnamefont {S.}~\bibnamefont {Das~Sarma}},\ }\href {\doibase
  10.1103/PhysRevB.93.184204} {\bibfield  {journal} {\bibinfo  {journal} {Phys.
  Rev. B}\ }\textbf {\bibinfo {volume} {93}},\ \bibinfo {pages} {184204}
  (\bibinfo {year} {2016})}\BibitemShut {NoStop}%
\bibitem [{\citenamefont {{Choudhury}}\ \emph {et~al.}(2018)\citenamefont
  {{Choudhury}}, \citenamefont {{Kim}},\ and\ \citenamefont
  {{Zhou}}}]{2018arXiv180705969C}%
  \BibitemOpen
  \bibfield  {author} {\bibinfo {author} {\bibfnamefont {S.}~\bibnamefont
  {{Choudhury}}}, \bibinfo {author} {\bibfnamefont {E.-a.}\ \bibnamefont
  {{Kim}}}, \ and\ \bibinfo {author} {\bibfnamefont {Q.}~\bibnamefont
  {{Zhou}}},\ }\href@noop {} {\bibfield  {journal} {\bibinfo  {journal} {ArXiv
  e-prints}\ } (\bibinfo {year} {2018})},\ \Eprint
  {http://arxiv.org/abs/1807.05969} {arXiv:1807.05969 [cond-mat.quant-gas]}
  \BibitemShut {NoStop}%
\bibitem [{\citenamefont {Schulz}\ \emph {et~al.}(2019)\citenamefont {Schulz},
  \citenamefont {Hooley}, \citenamefont {Moessner},\ and\ \citenamefont
  {Pollmann}}]{PhysRevLett.122.040606}%
  \BibitemOpen
  \bibfield  {author} {\bibinfo {author} {\bibfnamefont {M.}~\bibnamefont
  {Schulz}}, \bibinfo {author} {\bibfnamefont {C.~A.}\ \bibnamefont {Hooley}},
  \bibinfo {author} {\bibfnamefont {R.}~\bibnamefont {Moessner}}, \ and\
  \bibinfo {author} {\bibfnamefont {F.}~\bibnamefont {Pollmann}},\ }\href
  {\doibase 10.1103/PhysRevLett.122.040606} {\bibfield  {journal} {\bibinfo
  {journal} {Phys. Rev. Lett.}\ }\textbf {\bibinfo {volume} {122}},\ \bibinfo
  {pages} {040606} (\bibinfo {year} {2019})}\BibitemShut {NoStop}%
\bibitem [{\citenamefont {Kj\"all}\ \emph {et~al.}(2014)\citenamefont
  {Kj\"all}, \citenamefont {Bardarson},\ and\ \citenamefont
  {Pollmann}}]{PhysRevLett.113.107204}%
  \BibitemOpen
  \bibfield  {author} {\bibinfo {author} {\bibfnamefont {J.~A.}\ \bibnamefont
  {Kj\"all}}, \bibinfo {author} {\bibfnamefont {J.~H.}\ \bibnamefont
  {Bardarson}}, \ and\ \bibinfo {author} {\bibfnamefont {F.}~\bibnamefont
  {Pollmann}},\ }\href {\doibase 10.1103/PhysRevLett.113.107204} {\bibfield
  {journal} {\bibinfo  {journal} {Phys. Rev. Lett.}\ }\textbf {\bibinfo
  {volume} {113}},\ \bibinfo {pages} {107204} (\bibinfo {year}
  {2014})}\BibitemShut {NoStop}%
\bibitem [{\citenamefont {Geraedts}\ \emph {et~al.}(2017)\citenamefont
  {Geraedts}, \citenamefont {Regnault},\ and\ \citenamefont
  {Nandkishore}}]{Geraedts_2017}%
  \BibitemOpen
  \bibfield  {author} {\bibinfo {author} {\bibfnamefont {S.~D.}\ \bibnamefont
  {Geraedts}}, \bibinfo {author} {\bibfnamefont {N.}~\bibnamefont {Regnault}},
  \ and\ \bibinfo {author} {\bibfnamefont {R.~M.}\ \bibnamefont
  {Nandkishore}},\ }\href {\doibase 10.1088/1367-2630/aa93a5} {\bibfield
  {journal} {\bibinfo  {journal} {New Journal of Physics}\ }\textbf {\bibinfo
  {volume} {19}},\ \bibinfo {pages} {113021} (\bibinfo {year}
  {2017})}\BibitemShut {NoStop}%
\bibitem [{\citenamefont {Laflorencie}(2016)}]{LAFLORENCIE20161}%
  \BibitemOpen
  \bibfield  {author} {\bibinfo {author} {\bibfnamefont {N.}~\bibnamefont
  {Laflorencie}},\ }\href {\doibase
  https://doi.org/10.1016/j.physrep.2016.06.008} {\bibfield  {journal}
  {\bibinfo  {journal} {Physics Reports}\ }\textbf {\bibinfo {volume} {646}},\
  \bibinfo {pages} {1 } (\bibinfo {year} {2016})},\ \bibinfo {note} {quantum
  entanglement in condensed matter systems}\BibitemShut {NoStop}%
\bibitem [{\citenamefont {Bardarson}\ \emph {et~al.}(2012)\citenamefont
  {Bardarson}, \citenamefont {Pollmann},\ and\ \citenamefont
  {Moore}}]{PhysRevLett.109.017202}%
  \BibitemOpen
  \bibfield  {author} {\bibinfo {author} {\bibfnamefont {J.~H.}\ \bibnamefont
  {Bardarson}}, \bibinfo {author} {\bibfnamefont {F.}~\bibnamefont {Pollmann}},
  \ and\ \bibinfo {author} {\bibfnamefont {J.~E.}\ \bibnamefont {Moore}},\
  }\href {\doibase 10.1103/PhysRevLett.109.017202} {\bibfield  {journal}
  {\bibinfo  {journal} {Phys. Rev. Lett.}\ }\textbf {\bibinfo {volume} {109}},\
  \bibinfo {pages} {017202} (\bibinfo {year} {2012})}\BibitemShut {NoStop}%
\bibitem [{\citenamefont {Serbyn}\ \emph {et~al.}(2013)\citenamefont {Serbyn},
  \citenamefont {Papi\ifmmode~\acute{c}\else \'{c}\fi{}},\ and\ \citenamefont
  {Abanin}}]{PhysRevLett.110.260601}%
  \BibitemOpen
  \bibfield  {author} {\bibinfo {author} {\bibfnamefont {M.}~\bibnamefont
  {Serbyn}}, \bibinfo {author} {\bibfnamefont {Z.}~\bibnamefont
  {Papi\ifmmode~\acute{c}\else \'{c}\fi{}}}, \ and\ \bibinfo {author}
  {\bibfnamefont {D.~A.}\ \bibnamefont {Abanin}},\ }\href {\doibase
  10.1103/PhysRevLett.110.260601} {\bibfield  {journal} {\bibinfo  {journal}
  {Phys. Rev. Lett.}\ }\textbf {\bibinfo {volume} {110}},\ \bibinfo {pages}
  {260601} (\bibinfo {year} {2013})}\BibitemShut {NoStop}%
\bibitem [{\citenamefont {Khemani}\ \emph {et~al.}(2017)\citenamefont
  {Khemani}, \citenamefont {Lim}, \citenamefont {Sheng},\ and\ \citenamefont
  {Huse}}]{PhysRevX.7.021013}%
  \BibitemOpen
  \bibfield  {author} {\bibinfo {author} {\bibfnamefont {V.}~\bibnamefont
  {Khemani}}, \bibinfo {author} {\bibfnamefont {S.~P.}\ \bibnamefont {Lim}},
  \bibinfo {author} {\bibfnamefont {D.~N.}\ \bibnamefont {Sheng}}, \ and\
  \bibinfo {author} {\bibfnamefont {D.~A.}\ \bibnamefont {Huse}},\ }\href
  {\doibase 10.1103/PhysRevX.7.021013} {\bibfield  {journal} {\bibinfo
  {journal} {Phys. Rev. X}\ }\textbf {\bibinfo {volume} {7}},\ \bibinfo {pages}
  {021013} (\bibinfo {year} {2017})}\BibitemShut {NoStop}%
\bibitem [{\citenamefont {Richard}()}]{doi:10.1002/andp.201700042}%
  \BibitemOpen
  \bibfield  {author} {\bibinfo {author} {\bibfnamefont {B.}~\bibnamefont
  {Richard}},\ }\href {\doibase 10.1002/andp.201700042} {\bibfield  {journal}
  {\bibinfo  {journal} {Annalen der Physik}\ }\textbf {\bibinfo {volume}
  {529}},\ \bibinfo {pages} {1700042}},\ \Eprint
  {http://arxiv.org/abs/https://onlinelibrary.wiley.com/doi/pdf/10.1002/andp.201700042}
  {https://onlinelibrary.wiley.com/doi/pdf/10.1002/andp.201700042} \BibitemShut
  {NoStop}%
\bibitem [{\citenamefont {Oganesyan}\ and\ \citenamefont
  {Huse}(2007)}]{PhysRevB.75.155111}%
  \BibitemOpen
  \bibfield  {author} {\bibinfo {author} {\bibfnamefont {V.}~\bibnamefont
  {Oganesyan}}\ and\ \bibinfo {author} {\bibfnamefont {D.~A.}\ \bibnamefont
  {Huse}},\ }\href {\doibase 10.1103/PhysRevB.75.155111} {\bibfield  {journal}
  {\bibinfo  {journal} {Phys. Rev. B}\ }\textbf {\bibinfo {volume} {75}},\
  \bibinfo {pages} {155111} (\bibinfo {year} {2007})}\BibitemShut {NoStop}%
\bibitem [{\citenamefont {Maksymov}\ \emph {et~al.}(2019)\citenamefont
  {Maksymov}, \citenamefont {Sierant},\ and\ \citenamefont
  {Zakrzewski}}]{PhysRevB.99.224202}%
  \BibitemOpen
  \bibfield  {author} {\bibinfo {author} {\bibfnamefont {A.}~\bibnamefont
  {Maksymov}}, \bibinfo {author} {\bibfnamefont {P.}~\bibnamefont {Sierant}}, \
  and\ \bibinfo {author} {\bibfnamefont {J.}~\bibnamefont {Zakrzewski}},\
  }\href {\doibase 10.1103/PhysRevB.99.224202} {\bibfield  {journal} {\bibinfo
  {journal} {Phys. Rev. B}\ }\textbf {\bibinfo {volume} {99}},\ \bibinfo
  {pages} {224202} (\bibinfo {year} {2019})}\BibitemShut {NoStop}%
\bibitem [{\citenamefont {Rao}(2018)}]{0953-8984-30-39-395902}%
  \BibitemOpen
  \bibfield  {author} {\bibinfo {author} {\bibfnamefont {W.-J.}\ \bibnamefont
  {Rao}},\ }\href {http://stacks.iop.org/0953-8984/30/i=39/a=395902} {\bibfield
   {journal} {\bibinfo  {journal} {Journal of Physics: Condensed Matter}\
  }\textbf {\bibinfo {volume} {30}},\ \bibinfo {pages} {395902} (\bibinfo
  {year} {2018})}\BibitemShut {NoStop}%
\bibitem [{\citenamefont {Filippone}\ \emph {et~al.}(2016)\citenamefont
  {Filippone}, \citenamefont {Brouwer}, \citenamefont {Eisert},\ and\
  \citenamefont {von Oppen}}]{PhysRevB.94.201112}%
  \BibitemOpen
  \bibfield  {author} {\bibinfo {author} {\bibfnamefont {M.}~\bibnamefont
  {Filippone}}, \bibinfo {author} {\bibfnamefont {P.~W.}\ \bibnamefont
  {Brouwer}}, \bibinfo {author} {\bibfnamefont {J.}~\bibnamefont {Eisert}}, \
  and\ \bibinfo {author} {\bibfnamefont {F.}~\bibnamefont {von Oppen}},\ }\href
  {\doibase 10.1103/PhysRevB.94.201112} {\bibfield  {journal} {\bibinfo
  {journal} {Phys. Rev. B}\ }\textbf {\bibinfo {volume} {94}},\ \bibinfo
  {pages} {201112} (\bibinfo {year} {2016})}\BibitemShut {NoStop}%
\bibitem [{\citenamefont {De~Luca}\ \emph {et~al.}(2014)\citenamefont
  {De~Luca}, \citenamefont {Altshuler}, \citenamefont {Kravtsov},\ and\
  \citenamefont {Scardicchio}}]{PhysRevLett.113.046806}%
  \BibitemOpen
  \bibfield  {author} {\bibinfo {author} {\bibfnamefont {A.}~\bibnamefont
  {De~Luca}}, \bibinfo {author} {\bibfnamefont {B.~L.}\ \bibnamefont
  {Altshuler}}, \bibinfo {author} {\bibfnamefont {V.~E.}\ \bibnamefont
  {Kravtsov}}, \ and\ \bibinfo {author} {\bibfnamefont {A.}~\bibnamefont
  {Scardicchio}},\ }\href {\doibase 10.1103/PhysRevLett.113.046806} {\bibfield
  {journal} {\bibinfo  {journal} {Phys. Rev. Lett.}\ }\textbf {\bibinfo
  {volume} {113}},\ \bibinfo {pages} {046806} (\bibinfo {year}
  {2014})}\BibitemShut {NoStop}%
\bibitem [{\citenamefont {Bera}\ \emph {et~al.}(2015)\citenamefont {Bera},
  \citenamefont {Schomerus}, \citenamefont {Heidrich-Meisner},\ and\
  \citenamefont {Bardarson}}]{PhysRevLett.115.046603}%
  \BibitemOpen
  \bibfield  {author} {\bibinfo {author} {\bibfnamefont {S.}~\bibnamefont
  {Bera}}, \bibinfo {author} {\bibfnamefont {H.}~\bibnamefont {Schomerus}},
  \bibinfo {author} {\bibfnamefont {F.}~\bibnamefont {Heidrich-Meisner}}, \
  and\ \bibinfo {author} {\bibfnamefont {J.~H.}\ \bibnamefont {Bardarson}},\
  }\href {\doibase 10.1103/PhysRevLett.115.046603} {\bibfield  {journal}
  {\bibinfo  {journal} {Phys. Rev. Lett.}\ }\textbf {\bibinfo {volume} {115}},\
  \bibinfo {pages} {046603} (\bibinfo {year} {2015})}\BibitemShut {NoStop}%
\bibitem [{\citenamefont {Bera}\ \emph {et~al.}(2017)\citenamefont {Bera},
  \citenamefont {Martynec}, \citenamefont {Schomerus}, \citenamefont
  {Heidrich-Meisner},\ and\ \citenamefont
  {Bardarson}}]{doi:10.1002/andp.201600356}%
  \BibitemOpen
  \bibfield  {author} {\bibinfo {author} {\bibfnamefont {S.}~\bibnamefont
  {Bera}}, \bibinfo {author} {\bibfnamefont {T.}~\bibnamefont {Martynec}},
  \bibinfo {author} {\bibfnamefont {H.}~\bibnamefont {Schomerus}}, \bibinfo
  {author} {\bibfnamefont {F.}~\bibnamefont {Heidrich-Meisner}}, \ and\
  \bibinfo {author} {\bibfnamefont {J.~H.}\ \bibnamefont {Bardarson}},\ }\href
  {\doibase 10.1002/andp.201600356} {\bibfield  {journal} {\bibinfo  {journal}
  {Annalen der Physik}\ }\textbf {\bibinfo {volume} {529}},\ \bibinfo {pages}
  {1600356} (\bibinfo {year} {2017})},\ \Eprint
  {http://arxiv.org/abs/https://onlinelibrary.wiley.com/doi/pdf/10.1002/andp.201600356}
  {https://onlinelibrary.wiley.com/doi/pdf/10.1002/andp.201600356} \BibitemShut
  {NoStop}%
\bibitem [{\citenamefont {Lin}\ \emph {et~al.}(2018)\citenamefont {Lin},
  \citenamefont {Sbierski}, \citenamefont {Dorfner}, \citenamefont {Karrasch},\
  and\ \citenamefont {Heidrich-Meisner}}]{SciPostPhys.4.1.002}%
  \BibitemOpen
  \bibfield  {author} {\bibinfo {author} {\bibfnamefont {S.-H.}\ \bibnamefont
  {Lin}}, \bibinfo {author} {\bibfnamefont {B.}~\bibnamefont {Sbierski}},
  \bibinfo {author} {\bibfnamefont {F.}~\bibnamefont {Dorfner}}, \bibinfo
  {author} {\bibfnamefont {C.}~\bibnamefont {Karrasch}}, \ and\ \bibinfo
  {author} {\bibfnamefont {F.}~\bibnamefont {Heidrich-Meisner}},\ }\href
  {\doibase 10.21468/SciPostPhys.4.1.002} {\bibfield  {journal} {\bibinfo
  {journal} {SciPost Phys.}\ }\textbf {\bibinfo {volume} {4}},\ \bibinfo
  {pages} {002} (\bibinfo {year} {2018})}\BibitemShut {NoStop}%
\bibitem [{\citenamefont {Edwards}\ and\ \citenamefont
  {Thouless}(1972)}]{Edwards_1972}%
  \BibitemOpen
  \bibfield  {author} {\bibinfo {author} {\bibfnamefont {J.~T.}\ \bibnamefont
  {Edwards}}\ and\ \bibinfo {author} {\bibfnamefont {D.~J.}\ \bibnamefont
  {Thouless}},\ }\href {\doibase 10.1088/0022-3719/5/8/007} {\bibfield
  {journal} {\bibinfo  {journal} {Journal of Physics C: Solid State Physics}\
  }\textbf {\bibinfo {volume} {5}},\ \bibinfo {pages} {807} (\bibinfo {year}
  {1972})}\BibitemShut {NoStop}%
\bibitem [{\citenamefont {Serbyn}\ \emph {et~al.}(2017)\citenamefont {Serbyn},
  \citenamefont {Papi\ifmmode~\acute{c}\else \'{c}\fi{}},\ and\ \citenamefont
  {Abanin}}]{PhysRevB.96.104201}%
  \BibitemOpen
  \bibfield  {author} {\bibinfo {author} {\bibfnamefont {M.}~\bibnamefont
  {Serbyn}}, \bibinfo {author} {\bibfnamefont {Z.}~\bibnamefont
  {Papi\ifmmode~\acute{c}\else \'{c}\fi{}}}, \ and\ \bibinfo {author}
  {\bibfnamefont {D.~A.}\ \bibnamefont {Abanin}},\ }\href {\doibase
  10.1103/PhysRevB.96.104201} {\bibfield  {journal} {\bibinfo  {journal} {Phys.
  Rev. B}\ }\textbf {\bibinfo {volume} {96}},\ \bibinfo {pages} {104201}
  (\bibinfo {year} {2017})}\BibitemShut {NoStop}%
\bibitem [{\citenamefont {Khemani}\ \emph {et~al.}(2015)\citenamefont
  {Khemani}, \citenamefont {Nandkishore},\ and\ \citenamefont
  {Sondhi}}]{Khemani2015}%
  \BibitemOpen
  \bibfield  {author} {\bibinfo {author} {\bibfnamefont {V.}~\bibnamefont
  {Khemani}}, \bibinfo {author} {\bibfnamefont {R.}~\bibnamefont
  {Nandkishore}}, \ and\ \bibinfo {author} {\bibfnamefont {S.~L.}\ \bibnamefont
  {Sondhi}},\ }\href {\doibase 10.1038/nphys3344} {\bibfield  {journal}
  {\bibinfo  {journal} {Nature Physics}\ }\textbf {\bibinfo {volume} {11}},\
  \bibinfo {pages} {560} (\bibinfo {year} {2015})}\BibitemShut {NoStop}%
\bibitem [{\citenamefont {Luitz}\ \emph {et~al.}(2015)\citenamefont {Luitz},
  \citenamefont {Laflorencie},\ and\ \citenamefont
  {Alet}}]{PhysRevB.91.081103}%
  \BibitemOpen
  \bibfield  {author} {\bibinfo {author} {\bibfnamefont {D.~J.}\ \bibnamefont
  {Luitz}}, \bibinfo {author} {\bibfnamefont {N.}~\bibnamefont {Laflorencie}},
  \ and\ \bibinfo {author} {\bibfnamefont {F.}~\bibnamefont {Alet}},\ }\href
  {\doibase 10.1103/PhysRevB.91.081103} {\bibfield  {journal} {\bibinfo
  {journal} {Phys. Rev. B}\ }\textbf {\bibinfo {volume} {91}},\ \bibinfo
  {pages} {081103} (\bibinfo {year} {2015})}\BibitemShut {NoStop}%
\bibitem [{\citenamefont {Filho}\ \emph {et~al.}(2017)\citenamefont {Filho},
  \citenamefont {Saguia}, \citenamefont {Santos},\ and\ \citenamefont
  {Sarandy}}]{PhysRevB.96.014204}%
  \BibitemOpen
  \bibfield  {author} {\bibinfo {author} {\bibfnamefont {J.~L. C. d.~C.}\
  \bibnamefont {Filho}}, \bibinfo {author} {\bibfnamefont {A.}~\bibnamefont
  {Saguia}}, \bibinfo {author} {\bibfnamefont {L.~F.}\ \bibnamefont {Santos}},
  \ and\ \bibinfo {author} {\bibfnamefont {M.~S.}\ \bibnamefont {Sarandy}},\
  }\href {\doibase 10.1103/PhysRevB.96.014204} {\bibfield  {journal} {\bibinfo
  {journal} {Phys. Rev. B}\ }\textbf {\bibinfo {volume} {96}},\ \bibinfo
  {pages} {014204} (\bibinfo {year} {2017})}\BibitemShut {NoStop}%
\bibitem [{\citenamefont {Villalonga}\ \emph {et~al.}(2018)\citenamefont
  {Villalonga}, \citenamefont {Yu}, \citenamefont {Luitz},\ and\ \citenamefont
  {Clark}}]{PhysRevB.97.104406}%
  \BibitemOpen
  \bibfield  {author} {\bibinfo {author} {\bibfnamefont {B.}~\bibnamefont
  {Villalonga}}, \bibinfo {author} {\bibfnamefont {X.}~\bibnamefont {Yu}},
  \bibinfo {author} {\bibfnamefont {D.~J.}\ \bibnamefont {Luitz}}, \ and\
  \bibinfo {author} {\bibfnamefont {B.~K.}\ \bibnamefont {Clark}},\ }\href
  {\doibase 10.1103/PhysRevB.97.104406} {\bibfield  {journal} {\bibinfo
  {journal} {Phys. Rev. B}\ }\textbf {\bibinfo {volume} {97}},\ \bibinfo
  {pages} {104406} (\bibinfo {year} {2018})}\BibitemShut {NoStop}%
\bibitem [{\citenamefont {Peschel}(2003)}]{0305-4470-36-14-101}%
  \BibitemOpen
  \bibfield  {author} {\bibinfo {author} {\bibfnamefont {I.}~\bibnamefont
  {Peschel}},\ }\href {http://stacks.iop.org/0305-4470/36/i=14/a=101}
  {\bibfield  {journal} {\bibinfo  {journal} {Journal of Physics A:
  Mathematical and General}\ }\textbf {\bibinfo {volume} {36}},\ \bibinfo
  {pages} {L205} (\bibinfo {year} {2003})}\BibitemShut {NoStop}%
\bibitem [{\citenamefont {Pouranvari}\ and\ \citenamefont
  {Montakhab}(2017)}]{PhysRevB.96.045123}%
  \BibitemOpen
  \bibfield  {author} {\bibinfo {author} {\bibfnamefont {M.}~\bibnamefont
  {Pouranvari}}\ and\ \bibinfo {author} {\bibfnamefont {A.}~\bibnamefont
  {Montakhab}},\ }\href {\doibase 10.1103/PhysRevB.96.045123} {\bibfield
  {journal} {\bibinfo  {journal} {Phys. Rev. B}\ }\textbf {\bibinfo {volume}
  {96}},\ \bibinfo {pages} {045123} (\bibinfo {year} {2017})}\BibitemShut
  {NoStop}%
\end{thebibliography}%
\end{document}